\documentclass[]{aa} 

\usepackage{graphicx}
\usepackage{txfonts}
\usepackage{natbib}
\usepackage{longtable}
\usepackage{lscape}
\usepackage{color}

\providecommand{\sorthelp}[1]{}

\newcommand{\Planck}{{\it Planck }}
\newcommand{\Herschel}{{\it Herschel }}
\newcommand{\AKARI}{{\it AKARI }}
\newcommand{\ISO}{{\it ISO }}
\newcommand{\Spitzer}{{\it Spitzer }}

\begin{document}
 
\title{
{Modelling of dust emission of a filament in the Taurus molecular cloud}
}

\author{Mika Juvela\inst{1}
}

\institute{
Department of Physics, P.O.Box 64, FI-00014, University of Helsinki,
Finland, {\em mika.juvela@helsinki.fi}
}

\authorrunning{M. Juvela et al.}

\date{Received September 15, 1996; accepted March 16, 1997}

\abstract { 
Dust emission is an important tool in studies of star-forming clouds, as a
tracer of column density and indirectly via the dust evolution that is
connected to the history and physical conditions of the clouds. 
} 
{
We examine radiative transfer (RT) modelling of dust emission over an extended
cloud region, using a filament in the Taurus molecular cloud as an example. We
examine how well far-infrared observations can be used to determine both the
cloud and the dust properties.
}
{
Using different assumptions of the cloud shape, radiation field, and dust
properties, we fit RT models to Herschel observations of the Taurus filament.
Further comparisons are made with measurements of the near-infrared
extinction. The models are used to examine the degeneracies between the
different cloud parameters and the dust properties.
}
{
The results show significant dependence on the assumed cloud structure and the
spectral shape of the external radiation field. If these are constrained to
the most likely values, the observations can be explained only if the dust
far-infrared (FIR) opacity has increased by a factor of 2-3 relative to the
values in diffuse medium. However, a narrow range of FIR wavelengths provides
only weak evidence of the spatial variations in dust, even in the models
covering several square degrees of a molecular cloud.
}
{ 
The analysis of FIR dust emission is affected by several sources of
uncertainty. Further constraints are therefore needed from observations at
shorter wavelengths, especially regarding the trends in dust evolution.
}


\maketitle

\section{Introduction} \label{sect:intro}

A number of space-borne observatories have made observations over
near-infrared (NIR), mid-infrared (MIR), FIR, sub-millimetre, and millimetre
wavelengths, and over extended regions of the interstellar medium (ISM):
\Planck at $\lambda\ge 350\mu{\rm m}$ \citep{Tauber2010}, \Herschel at
$\lambda$=70-500\,$\mu{\rm m}$ \citep{Pilbratt2010}, \ISO at 2.4-250\,$\mu$m
\citep{Kessler1996}, \AKARI at $\lambda$=1.7-180\,$\mu$m\citep{Murakami2007},
\Spitzer at 3.6-160\,$\mu$m \citep{Werner2004}, and WISE at 3-22\,$\mu$m
\citep{Wright2010}. 
This means that many nearby clouds have been observed with full wavelength
coverage of the observable dust emission. The data have shown that the
emission properties of interstellar dust are not constant at Galactic scales
\citep{Planck2014_allsky_model} and also change drastically between diffuse
and molecular regions \citep{Lagache1998, Cambresy2001} and even at smaller
scales. One of the effects is the enhanced FIR/sub-millimetre opacity of dense
molecular clouds, which is believed to be connected to the formation of ice
mantles, increased grain sizes, and the formation of dust aggregates in the
densest and coldest parts of the molecular clouds. A strong effect was
observed in a filament in Taurus already in \citep{Stepnik2003}, and this is
now known to be a common phenomenon of the densest parts of molecular clouds
\citep{Roy2013, GCC-V}. 

The changes in dust are reflected not only in the absolute opacity but also 
in the opacity spectral index $\beta$, which gets typically has values in the
range $\beta$=1.5-2.0, depending on both the observed source and the
wavelengths. Balloon-borne telescopes \citep{Bernard1999, Dupac2003,
Desert2008, Paradis2009} and the analysis of \Herschel and \Planck data
\citep{Paradis2010, Paradis2012, Planck2014_XI, PlanckXXIX2016} revealed
variations over wide sky areas. Ground-based instruments have extended the
studies to millimetre wavelengths, with resolution comparable to the
space-borne FIR data. There is evidence of FIR spectrum getting steeper
towards dense clumps and cores \citep{GCC-VI,Scibelli2023}, but the spectrum
flattens ($\beta$ decreases) at longer wavelengths \citep{Paradis2012_500um,
Planck2014_XI}. \citet{Mason2020} found in Orion significant 3\,mm and 1\,cm
excess over the $\beta=1.7$ spectrum appropriate in the FIR regime, a possible
sign of the presence of extremely large grains.

Dust observations are often analysed via fits to the spectral energy
distribution (SED). However, the warmest dust component within the telescope
beam, including the line-of-sight (LOS) variations, has the largest effect on
the SED. Therefore, the assumption of single temperature results in an
overestimation of the (mass-averaged) dust temperature \citep{Shetty2009a,
Juvela2012_Tmix} and the dust column densities are underestimated. With
observations at more frequencies, the data can be modelled as the sum of
several temperature components or a general distribution of dust temperatures
\citep{Juvela2023_MBB}. The Abel transform has been applied to the analysis of
cores \citep{Roy2014, Bracco2017}, making use of the strong constraint of the
spherical symmetry of the source, which is then mostly applicable for isolated
cores. More generally, methods like point process mapping
\citep[PPMAP;][]{Marsh2015} decompose the spectrum along each line of sight
into a sum of temperature components. PPMAP has been applied also to extended
cloud regions, including cloud filaments in the Taurus molecular cloud and
elsewhere \cite{Howard2019, Howard2021}.

The simultaneous determination of the temperature and $\beta$ suffers from 
degeneracy, which makes the results sensitive to noise and any sources of
systematic errors \citep{Shetty2009b, Veneziani2010, Juvela2012_bananasplit}.
This is only partly remedied by the use of hierarchical Bayesian modelling
\citep{Kelly2012, Juvela2013_TBmethods}, where the additional constraint comes
from the assumed similarity in the ($T$, $\beta$) values in the observational
set. With observations over a sufficiently wide wavelength range and with a
sufficiently high signal-to-noise ratio (SN), it should be possible to extract
some information also of spectral index variations and even the LOS
temperature variations. Both are needed for accurate optical depth estimates
and better column density and mass estimates.

What the empirical SED-fitting methods lack is the requirement of physical
self-consistency between the radiation field, the cloud structure, and the
dust properties. Short wavelengths (from ultraviolet to NIR) determine how
much energy dust absorbs and how this heating varies over the source. The
energy is re-emitted at longer wavelengths, mainly in the FIR. The FIR dust
properties affect the final dust temperature and, especially via $\beta$, the
shape of the observed SEDs. Full self-consistency between the different
wavelengths and different source regions is enforced only in a complete RT
model. This should be a useful additional constraint when observations are
used to study dust properties and especially their spatial variations.

The advantages of full modelling are not always self-evident. When the whole
source volume is interconnected via radiation transport and the density field
is complex, it is very difficult to find a model that reproduces all
observations within the observational uncertainties. The predictions of a
model that fits observations only approximately are naturally less reliable.
In contrast, in the simple SED fits the parameters ($T$ and possibly $\beta$)
can be tuned for each pixel separately. Therefore, the observations are
matched well at each position, and the main uncertainties are caused by the
more naive assumptions underlying the analysis (e.g. a single temperature).

Even if a RT model does not fit the observations perfectly, the fit residuals
can still provide valuable information on where the physical conditions or the
dust properties differ from the model assumptions. The complexity of RT models
(LOS cloud structure, spectrum and intensity of the external radiation field,
positions and properties of potential embedded radiation sources, etc.) makes
it difficult to cover all feasible parameter combinations. Thus, although a
good fit might be reached with some RT model, the mapping of all potentially
correct models (and the associated uncertainties) is hardly possible.

In this paper we construct large-scale RT models for the LDN~1506 (B212-B215)
filament in the Taurus molecule cloud.  The small distance
\citep[$\sim$129\,pc for B215;][]{Galli2019} and the absence significant local
heating makes the region relatively easy to model. The models cover an area of
$1.8 \times 1.8$ degrees, and they are optimised to reproduce the FIR
observations over the full field. The FIR emission of individual filament
cross sections was already modelled in \citet{Ysard2013}, suggesting clear
dust evolution and increased FIR dust emissivity in the filament. Rather than
trying to develop a definite model for the region and its dust property
variations, we examine to what extent the FIR observations are able to
constrain the dust properties. We compare models with different dust
properties and investigate other factors, such as the 3D cloud density field
and the external radiation field, that could confuse the evidence for dust
evolution. In these models, the model parameters (apart from possible
density-dependent changes in dust properties) cannot be tuned locally, so a
perfect match to the observations is not to be expected. However, if the
modelling were successful (and preferably without significant degeneracies),
it could provide simultaneously valuable information on the dust properties,
the cloud mass and mass distribution, and the radiation field.

The contents of the paper are the following. We present the observational data
in Sect.~\ref{sect:observations} and the methods of RT modelling and employed
dust models are presented in Sect.\ref{sect:methods}. The results from the
optimised RT models with basic dust models are presented in
Sect.~\ref{sect:results}. We discuss the results in
Sect.~\ref{sect:discussion}, where we also investigate further the dust
property changes that are needed to explain the Taurus observations. The final
conclusions are listed in Sect.~\ref{sect:conclusions}.

\begin{figure}
\includegraphics[width=8.8cm]{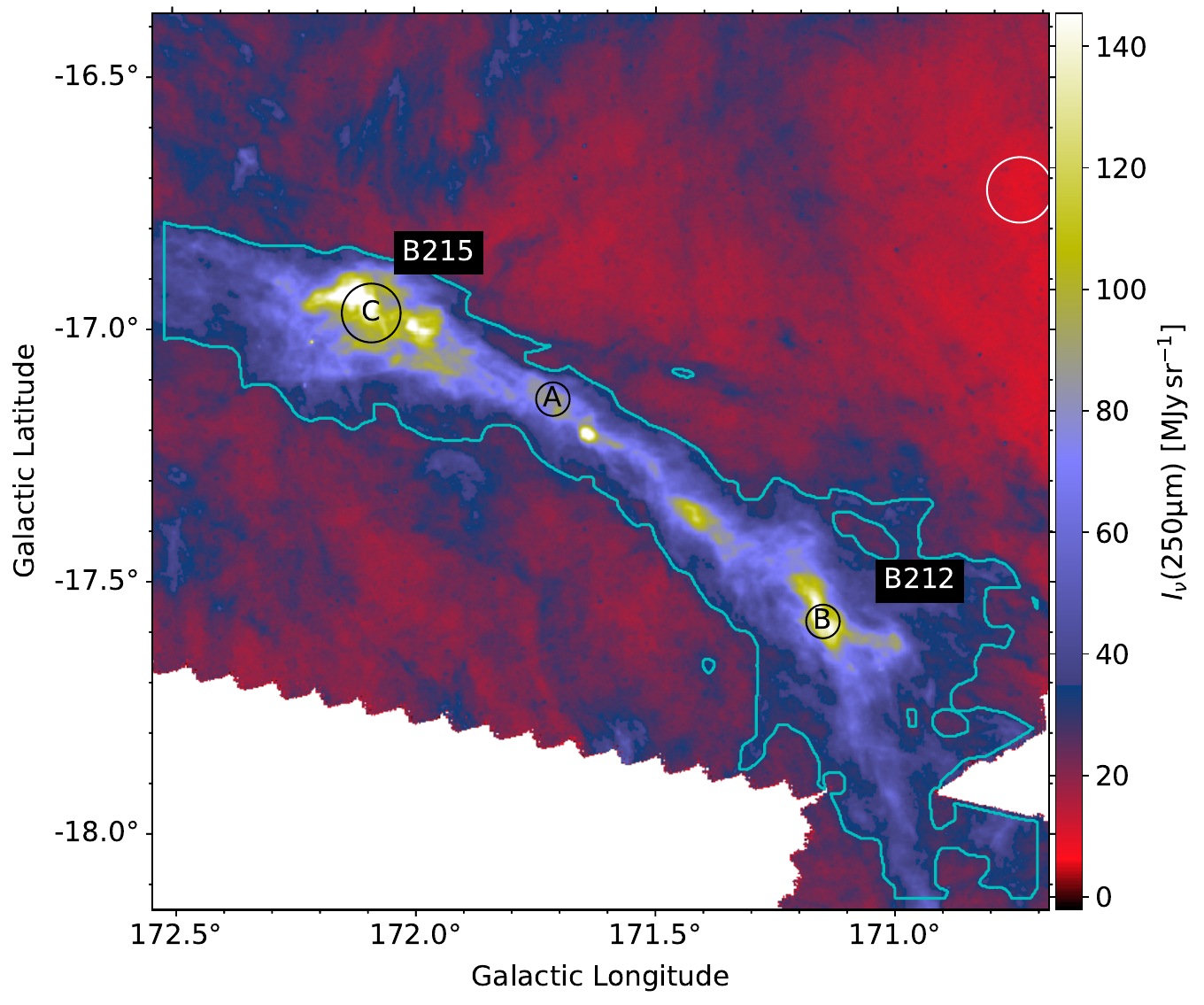}
\caption{
\Herschel 250\,$\mu$m map of the Taurus filament. The model fits are compared
mainly to the data inside the cyan contour (``filament region''), which
corresponds to 10\,MJy\,sr$^{-1}$ in the background-subtracted
350\,$\mu$m map. The circles labelled A-C correspond to selected small regions, in
order of increasing column density, that are used in the model
comparisons. The white circle shows the area used for background subtraction.
}
\label{fig:plot_map}
\end{figure}

\section{Observational data} \label{sect:observations}

The cloud models are optimised based on 250-500\,$\mu$m surface brightness
observations. Measurements of NIR extinction and 160\,$\mu$m emission are used
afterwards to check, how well the models predict these shorter-wavelength
data.

\subsection{Surface brightness maps}

The 250\,$\mu$m, 350\,$\mu$m, and 500\,$\mu$m FIR surface brightness maps were
all observed with the \Herschel spectral and photometric imaging receiver
(SPIRE) instrument, and the relative accuracy of the observations is expected
to be better than
2\%\footnote{https://www.cosmos.esa.int/web/herschel/spire-overview gives 4\%
accuracy for the absolute and 1.5\% for the relative SPIRE calibration}. The
Taurus SPIRE observations we made as part of the Gould Belt project (PI Ph.
Andre; parallel mode maps, OBS ID 1342204860 and 1342204861). The angular
resolution of the observations is about 18, 26, and 37 arcsec for the three
bands in the order of increasing wavelength. For the modelling, the
250\,$\mu$m and 350\,$\mu$m were degraded to correspond to the Gaussian beam
with the full width at half maximum (FWHM) equal to 30$\arcsec$. The
500\,$\mu$m data were similarly degraded to FWHM=40$\arcsec$. The lower
resolution relaxes the requirements for the spatial resolution of the RT
models and reduces the uncertainties connected to the beam shapes. All maps
were resampled onto the same 10$\arcsec$ pixels in galactic coordinates. 

We subtracted from the maps the background emission that was estimated as the
mean value within 3.9 arcmin of the position ($l$,$b$)=(170.747\,deg,
-16.724\,deg) (Fig.~\ref{fig:plot_map}). The data were colour corrected for
modified blackbody (MBB) spectra $\propto B_{\nu}(T)\times \nu^{\beta}$, where
the temperatures were obtained from MBB fits at 40$\arcsec$ resolution and
assuming $\beta=1.8$. The colour corrections of the SPIRE channels are small
(a couple of percent) and change only slowly as functions of temperature and
spectral index. The 250\,$\mu$m map is shown in Fig.~\ref{fig:plot_map}.

We also use \Herschel 160\,$\mu$m maps that were observed with the
photodetector array camera and spectrometer (PACS) instrument (PACS photometry
maps, OBS ID 1342227304 and 1342227305). The data are convolved down to
30$\arcsec$ resolution and colour corrected for the same SED shape as the
SPIRE channels. We assume for the PACS data a 4\% uncertainty relative to the
SPIRE data.

\subsection{Extinction of background stars} \label{sect:extinction}

\begin{figure}
\includegraphics[width=8.8cm]{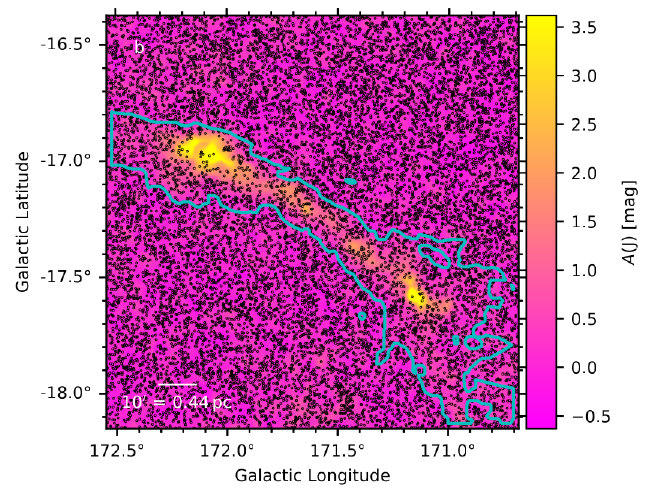}
\caption{
The 2MASS stars (black dots) plotted on the NICER $A({\rm J})$ extinction map.
The nominal resolution of the extinction  map is 2 arcmin.
}
\label{fig:A_extinction}
\end{figure}

NIR extinction can provide a good discriminator for different models that
match the same FIR observations \citep{Juvela_L1642}. To estimate the NIR
extinction, we use stars from the 2MASS survey \citep{Skrutskie2006}.
Figure~\ref{fig:A_extinction} shows the J-band extinction map calculated with
the NICER method \citep{Lombardi2001} as well as the distribution of the
individual stars. The area contains some 14 700 stars, 2364 of which are
inside the cyan contour in Fig.~\ref{fig:plot_map}, which we refer to as the
filament region. The extinction map is made using the $R_{\rm V}$=4 extinction
law \citep{Cardelli1989}. Due to the high density of the filament region, we
adopt a value above the normal ISM value of $R_{\rm V}$=3.1, but the $R_{\rm
V}$ difference has a minimal effect at NIR wavelengths
\citep{Cardelli1989,MartinWhittet1990,Hensley2023}.

The resolution of the extinction map (2$\arcmin$ in
Fig.~\ref{fig:A_extinction}) is restricted by the number of background stars,
and the stellar density decreases towards the densest regions. This results in
higher uncertainty in $A({\rm J})$ towards density peaks and systematic errors
in regions of extinction gradients. Some or even most of the bias can be
eliminated statistically \citet{Lombardi2009,Lombardi2018}. However, in the
following analysis, we will use the individual stars rather than the
continuous extinction map. The stars probe the extinction towards discrete
positions, without the uncertainty of spatial interpolation.  In spite of
photometric errors and uncertainty of the intrinsic colours of the individual
stars, the number of stars is sufficient for reliable comparisons with the
predictions of the fitted RT models. The latter have 30$\arcsec$ resolution
and, by construction, no structure at scales below 10$\arcsec$.

\section{Methods}  \label{sect:methods}

The modelling includes the construction of the initial model setup, including
the selection of a dust model, and the optimisation of the model against FIR
observations. Here the word ``model'' refers to the combination of the chosen
density field, the description of the radiation sources, and the specification
of the dust model.

\subsection{Density field}  \label{sect:density}

The initial density field was constructed based on the observed 250\,$\mu$m
surface brightness, combined with an assumption of the line-of-sight (LOS)
density profile. The absolute values or the variations in the initial mass
distribution over the plane of the sky (POS) are not import because the final
values will result from the the model optimisation. On the other hand, the
assumed LOS cloud structure is an important parameter. A larger extent means
that the medium receives more radiation, leading to smaller temperature
gradients and higher surface brightness per column density. We used two
options for the LOS structure. The first one is a Gaussian profile that is
fully determined by the FWHM of the LOS density distribution. We use three
values $FWHM$=0.2, 0.5, and 0.9\,pc. Here $FWHM$=0.2\,pc is closest to the
expected size of interstellar filaments, while the larger values might account
for more extended cloud structures or a sheet-like geometry along the LOS
direction.
Alternatively, we use Plummer-type density profiles \citep{Arzoumanian2011}.
Cross sections of the column density of the main filament were fitted with the
Plummer model, which was then converted to the LOS density profile under the
assumption of rotational symmetry. The parameters of the Plummer fit are the
central density, the size of the central flat region, and the asymptotic
powerlaw index $p$. The parameters were allowed to vary along the filament,
but the $p$ parameter was limited to the range 1.7$-$3.5.

For the RT calculations, the densities were discretised onto a
three-dimensional hierarchical octree grid. The modelled area is $106.7 \times
106.7$ arcmin in size ($\sim$3.2 square degrees), and with a pixel size of
10$\arcsec$ corresponds to $640\times640$ resolution element. The octree grid
has a root grid of 80$^3$ cells. With three levels of refinement, the smallest
cell size thus corresponds to the 10$\arcsec$ pixel size, and all maps of the
models have the same $640\times640$ pixels as the observations. The refinement
was based on volume density. By using the hierarchical discretisation, the
total number of volume elements could be kept around 15 million. During
the optimisation, the density field is modified based on the observed and
model-predicted 350\,$\mu$m surface brightness maps, and the ratio in each map
pixel is used to scale the densities in all cells along the same LOS. The
correction changes the model column densities but does not change the original
LOS density profile.

\subsection{Radiation field}  \label{sect:radiation}

The dust is in the simulations heated by anisotropic background radiation. The
angular distribution of the sky brightness is taken from COBE DIRBE allsky
maps \citep{Boggess1992}, where the closest band is also used for frequencies
outside the observed range. DIRBE is used only for the angular distribution,
and the values were rescaled to match the level and spectral shape of the
\citet{Mathis1983} model of the interstellar radiation field (ISRF). The
strength of this external field is in RT models left as a free parameter, with
the same multiplier for all frequencies. This scalar scaling parameter is
optimised based on the observed and the model-predicted average surface
brightness ratios $I_{\nu}(250\,\mu{\rm m}/I_{\nu}(500\,\mu{\rm m}$ that are
calculated over the filament region (inside the cyan contour in
Fig.~\ref{fig:plot_map}). The more diffuse regions are excluded, also because
they could be subjected to different radiation fields along the LOS and could
include more extended structures than in the model that is limited to a finite
LOS depth.

Because the modelled region is embedded in an extended molecular cloud, we
consider alternative radiation fields that have suffered $A_{\rm V}$=0-2
magnitudes of extinction due to cloud layers outside the modelled volume. The
field is attenuated according to the extinction curve of the
\citet{Compiegne2011} dust model. Extinction will remove more energy from the
shortest wavelengths. The incoming radiation is then effectively moved to
longer wavelengths where it suffers less attenuation, leading to smaller
temperature gradients inside the model volume. The effect is therefore
qualitatively similar to the effect of a larger LOS extent of the model cloud.

Based on \Planck dust optical depth maps, the LOS extinction around inside the
modelled areas and outside the main filaments is $A_{\rm V}\sim1$\,mag. The
scaling from 353\,GHz includes large uncertainty, and the visual extinction
could be smaller if one assumed the FIR opacity to be above the values found
in diffuse regions of the Milky Way \citep{Boulanger1996,planck2013-p06b}.
Since the external layer would correspond to half of the full LOS extinction
(and mutual shadowing provided by the dense filaments is quite small), the
external layer is expected to correspond to less than $A_{\rm V}$=1\,mag.

The dust heating should be caused mainly by the stellar radiation field,
which has some uncertainty in its level and spectral shape. Short-wavelength dust
emission, such as from polycyclic aromatic hydrocarbons (PAHs), could also
contribute to the dust heating at high optical depths, when the
short-wavelength part of the ISRF is strongly attenuated. Also this would be
part of the general ISRF rather than of local origin, since the reduced
abundance and excitation of PAHs reduces their emission inside dense
molecular clouds. The combination of the ISRF scaling factor and the assumed
external attenuation is effectively changing the balance between the heating
by UV-optical and NIR-MIR radiation, whether the latter is due to PAHs or 
other sources. The contribution of even longer wavelengths (e.g. of the cosmic
microwave background) is insignificant at the optical depths of the Taurus
filaments.

We include in some models also point sources in order to test the effects of
potential embedded sources. The luminosity of each source is left as a free
parameter (reported in units of the solar luminosity) and the emission is 
modelled as pure blackbody radiation. The source temperature is a again
relevant parameter, because it affects the wavelength distribution of the
heating radiation and how localised the effect of each point source remains.
Because of the finite resolution of the models, the temperature assigned to
the sources corresponds to the escaped radiation at some 1500\,au distance
from the source (the minimum cell size in the models), not the intrinsic
emission of an embedded (young) stellar objects. The source luminosities are
optimised based on the surface brightness at 20-60$\arcsec$ distance of each
point source. The direct LOS to the source is excluded, because
that is more dependent on coarse resolution of the RT model.

\subsection{Dust models} \label{sect:dust}

One of the main goals of the paper is to compare how well observations can be
matched with different assumptions of the dust properties. We use three basic
dust models, the \citet{Compiegne2011} model (in the following COM), and the
core–mantle–mantle grains (CMM) and aggregates with ice mantles (AMMI) from
the THEMIS dust model \citep{Jones2013,Ysard2016}. We also include two ad hoc
variations of the CMM model. In CMM-1 the opacity spectral index $\beta$ is
artificially reduced by 0.3 units for all wavelengths $\lambda>70\,\mu$m,
while keeping the 250\,$\mu$m opacity unchanged. In CMM-2 the
$\lambda>70\,\mu$m opacities are increased by 50\%, with no change in $\beta$.

\begin{figure}
\begin{center}
\includegraphics[width=8cm]{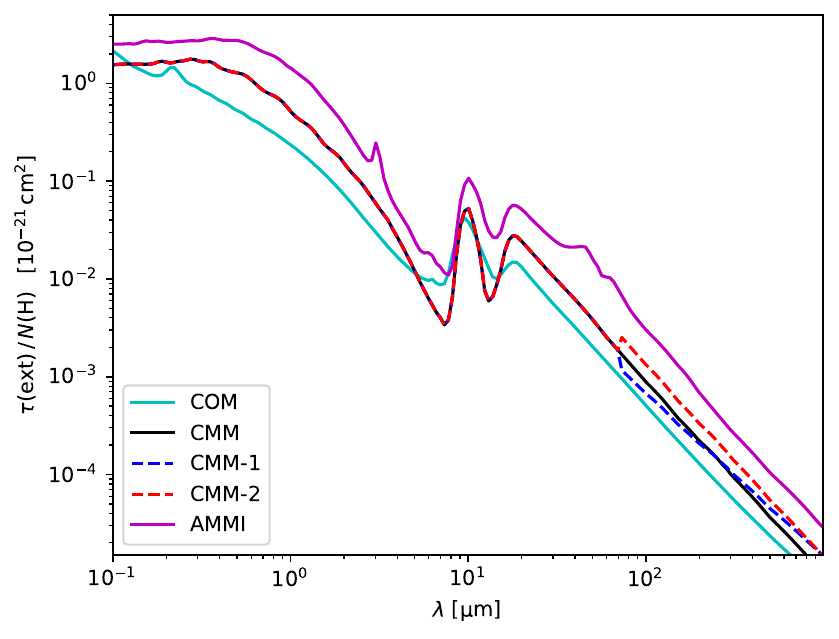}
\end{center}
\caption{
Extinction curves of the adopted dust models.  Below 70\,$\mu$m the CMM-1 and
CMM-2 variants are identical to CMM. 
}
\label{fig:plot_dusts}
\end{figure}

The extinction curves (optical depth per Hydrogen atom) of the dust models are
shown in Fig.~\ref{fig:plot_dusts}. It is worth noting that the absolute level
of the extinction curve will have no effect on the quality of the FIR fits nor
the predictions for the NIR extinction. It will affect on the mass estimates
(the conversion from optical depth to mass including division with opacity
$\kappa_{\nu}$). On the other hand, the ratio between the optical and NIR
wavelengths (where dust absorbs energy) and the FIR wavelengths (where energy
is re-emitted) is important for the FIR fits and especially for the predicted
NIR extinction values.

In most runs, the same dust properties are used throughout the model volume.
We test later also some models with a smooth transition from one set of dust
properties at low densities to another at high densities. The transition is
implemented with the help of fractional abundances
\begin{equation}
\chi = \frac{1}{2} - \frac{1}{2}
\tanh [4.0 \times (\log_{\rm 10} n_{\rm H}-\log_{\rm 10} n_{\rm 0} )].
\label{eq:abu}
\end{equation}
Thus, the model includes two dust components with fractional abundances $\chi$
and $1-\chi$, where the values depend on the local volume density $n({\rm H})$
and the pre-selected density threshold $n_{\rm 0}({\rm H})$.

\subsection{Radiative transfer calculations} \label{sect:RT}

The radiative transfer calculations were carried out with the SOC Monte Carlo
program \citep{Juvela2019_SOC}, using the volume discretisation described in
Sect.~\ref{sect:density}. Thanks to the use of hierarchical grids, the model
resolution was in dense regions 10$\arcsec$, better than for the observations,
while the run times were still manageable. To keep the Monte Carlo noise at
$\la$1\% level (well below that of observations), we used $\sim 10^7$ photon
packages per frequency for both the external field and each of the point
sources. For point sources this is somewhat excessive, because the influence
of each source is limited to a small region. SOC can handle spatial variations
in the dust properties if, as described in Sect.~\ref{sect:dust}, this is
described using abundance variations for a finite number of dust components. A
single RT run took between $\sim$10 seconds and a couple of minutes, depending
on the number of point sources and dust components. Because the analysis was
limited to emission at long wavelengths, $\lambda \ge 160\,\mu{\rm m}$, the
stochastic heating (relevant only for small grains) was not solved, and the
dust was assumed to be in equilibrium with the radiation field.

\subsection{Modified blackbody fits} \label{sect:MBB}

The optical depths of the RT models will be compared to the values obtained by
fitting the SEDs with MBB functions. In the case of a single temperature
component and optically thin emission, the optical depth is
\begin{equation}
\tau_{\nu} =  I_{\nu} / B_{\nu}(T_{\rm d}),
\end{equation}
where the dust temperature $T_{\rm d}$ is obtained by fitting the
multi-frequency observations with a MBB function,
\begin{equation}
I_{\nu} \propto  B_{\nu}(T_{\rm d}) \times \nu^{\beta}.
\end{equation}
The dust opacity is here assumed to follow a powerlaw, $\kappa_{\nu} \propto
\nu^{\beta}$. For this analysis, the surface brightness data are first
convolved to the same resolution, which makes it possible to do the
calculations for each map pixel separately.

\citet{Juvela2023_MBB} discussed alternative SED fits, where the dust
temperatures is assumed to follow a normal distribution and both the mean
temperature and the width of this temperature distribution are free
parameters. The fit was done with Markov chain Monte Carlo (MCMC) methods and
the angular resolution of the maps at different frequencies was not required
to be the same. However, in this paper we use the observations at a common
angular resolution, the same as in the case of the single-temperature fits.

\section{Results} \label{sect:results}

We present below results from the fitting of alternative models to the FIR
observations of the Taurus B212-B215 (L1506) filament. Results are shown for
single-dust models (Sect.~\ref{sect:L1506_single}), testing separately the
potential effects of embedded sources (Sect.~\ref{sect:L1506_stars}), before
first experiments with spatial variations in the dust properties
(Sect.~\ref{sect:L1506_twin}).

\subsection{Single-dust models} \label{sect:L1506_single}

We fitted the observations with RT models with COM, CMM, CMM-1, CMM-2, and
AMMI dust models (Sect.~\ref{sect:dust}), using constant dust properties over
the model volume. The optimised parameters are the column densities (adjusted
pixel by pixel) and the strength of the external radiation field (a scalar
parameter). Since the strength of the radiation field is optimised using the
average SED in the filament region, the fit concentrates on matching the
average SED shape in that area. We tested alternative models that differ
regarding the LOS density profile (Sect.~\ref{sect:density}) and the external
radiation field (Sect.~\ref{sect:radiation}).

\begin{figure}
\includegraphics[width=8.8cm]{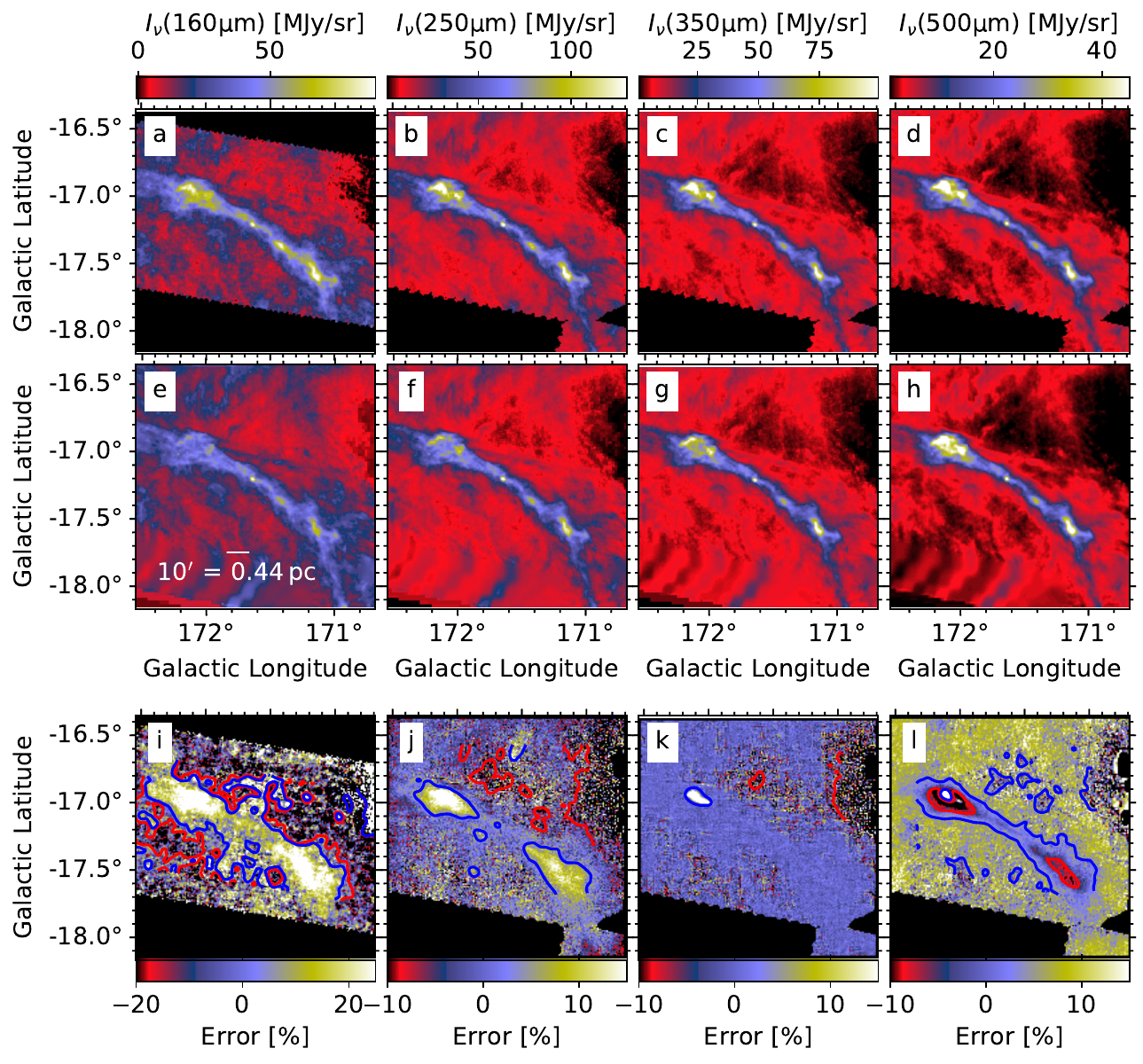}
\caption{
Fit using the CMM dust model, $FWHM=0.5$\,pc, and $A_{\rm V}=0$\,mag. The
first row of frames shows the observations, the second row the predictions of
the fitted RT model, and the third row the fit residuals. The actual fit used
only 250-500\,$\mu$m data. The data are plotted at the resolution of the used
observations, but the contours in the bottom frames show the $\pm$4\% error
levels (red and blue contours, respectively) when, for clarity, the data have
been smoothed to 3 arcmin resolution.
}
\label{fig:plot_basic_fit}
\end{figure}

Figure~\ref{fig:plot_basic_fit} presents as an example of a fit carried out
with the CMM dust model, a Gaussian LOS density profile with $FWHM$=0.5\,pc,
and no external attenuation of the ISRF. The upper frames show the observed
160-500\,$\mu$m maps. The second row shows the surface brightness predictions
from the RT model fitted to the 250-500\,$\mu$m observations. The 160\,$\mu$m
map is therefore a prediction to a wavelength outside the fitted range. In the
lower left map corner, the RT model is extrapolated outside the SPIRE coverage
in order to reduce edge effects close to the filament. 

The last row of frames in Fig.~\ref{fig:plot_basic_fit} shows the fit
residuals (percentage of the surface brightness). Because the column density
is adjusted separately for each map pixel and using the observed and
model-predicted 350\,$\mu$m maps, the 350\,$\mu$m residuals should always be
close to zero. However, Fig.\ref{fig:plot_basic_fit}k shows one example how
the fit can fail, in this case near the region C (the north-eastern core). The
level of the external radiation field is set based on the average signals in
the filament area. However, with adopted radiation field SED and the dust
properties, the model is unable to produce sufficiently high surface
brightness around one position (region C). This could also happen if the core
had an embedded radiation source, which is not part of the model. The positive
350\,$\mu$m residuals indicate that the optimisation has at this one position
increased the column density to the set upper limit (ten times the initial
analytical column density estimate), and the fit in that region is incorrect.

In Fig.~\ref{fig:plot_basic_fit}, the 250\,$\mu$m and 500\,$\mu$m residuals
are up to 10\% level and thus significant compared to the relative accuracy
between the SPIRE bands. At the extrapolated 160\,$\mu$m wavelength the errors
increase beyond 20\%. The model tends to be too cold along the central
filament, leading to residuals (observation minus model prediction) that are
positive at 250\,$\mu$m and negative at 500\,$\mu$m. Because the model tries
to match the average 250\,$\mu$m/500\,$\mu$m ratio over the whole filament
region, the errors are always relative. Thus, for the optimised radiation
field, the emission is too cold in the inner and too warm in the outer parts
of the filament. It also means that the temperature gradients are stronger in
the model than in the real cloud, and the model probably has too high optical
depth at the short wavelengths that are responsible for the dust heating.

\begin{figure}
\includegraphics[width=8.8cm]{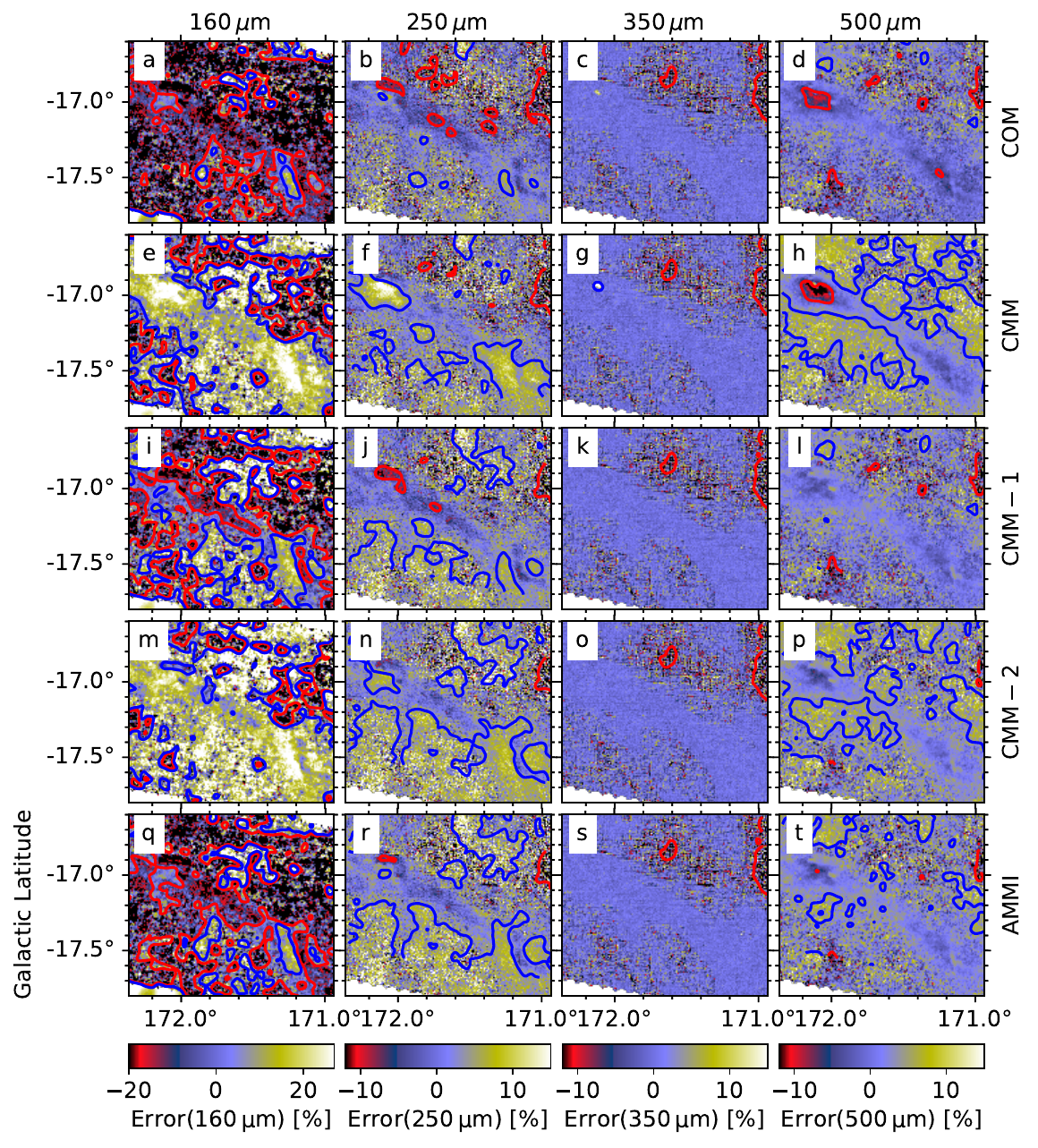}
\caption{Comparison of fit residuals in fits of single-dust models. The cloud
LOS extent is $FWHM=0.5$\,pc and the attenuation of the external radiation
field corresponds to $A_{\rm V}=1$\,mag. The rows correspond to the
dust models COM, CMM, CMM-1, CMM-2, and AMMI, respectively.
}
\label{fig:plot_residuals_3}
\end{figure}

Figure~\ref{fig:plot_residuals_3} shows fits where the external radiation
field is attenuated by $A_{\rm V}=1^{\rm mag}$. Because the level of the
radiation field is a free parameter, the effect is only to change the shape of
this spectrum. As the external extinction removes energy preferentially from
the short wavelengths, the model is effectively optically thinner for the
remaining radiation, and this should reduce the temperature variations inside
the model. Compared to the previous $A_{\rm V}=0^{\rm mag}$ case, the fits
with the CMM dust model show some improvement, and the positive 350\,$\mu$m
residuals cover a smaller area. The 250\,$\mu$m errors are almost down to the
4\% level. Interestingly, the 500\,$\mu$m data, which should be less sensitive
to temperature variations, show more significant variations in the residuals
across the filament.

Figure~\ref{fig:plot_residuals_3} also shows results for two ad hoc
modifications of the CMM dust model. CMM-1 has a 0.3 units lower $\beta$ with
no change in the 250\,$\mu$m opacity, and CMM-2 retains the original $\beta$
but with 50\% higher FIR opacity. Both CMM-1 and CMM-2 result in some
improvement in the fits. The 250-500\,$\mu$m fit is almost perfect with CMM-1
(with lower $\beta$), but the extrapolation to 160\,$\mu$m now overestimates
rather than underestimates the emission. The CMM-2 (with higher absolute FIR
opacity) is close to the original CMM, but with lower errors especially at
500\,$\mu$m.

The figure shows results also for two other dust models, the COM model of
diffuse medium and, with its ice mantles, AMMI in principle appropriate for
the densest parts of molecular clouds. Interestingly, both result in very
similar fit quality, the match to the 500\,$\mu$m data being better than with
CMM. Unlike the CMM model that tends to underestimate the 160\,$\mu$m
intensity within the filament, both COM and AMMI lead to some overestimation
at this wavelength.

\begin{figure}
\includegraphics[width=8.8cm]{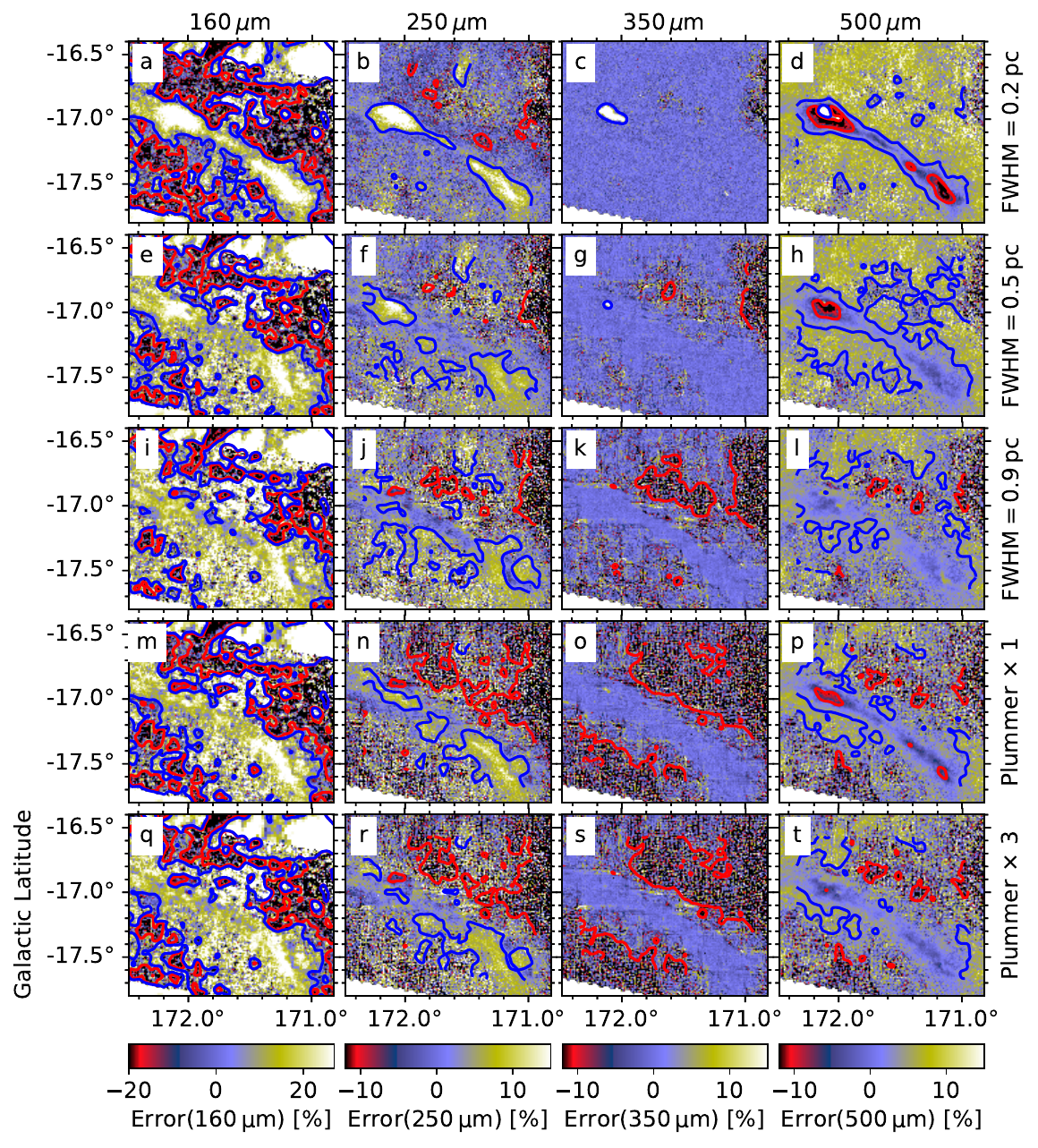}
\caption{Comparison of fit residuals for models of different cloud LOS extent.
The upper three rows correspond to Gaussian LOS density profiles with FWHM
equal to 0.2, 0.5, and 0.9\,pc, respectively. On the fourth row the LOS
profile is based on Plummer fits to the filament profile, and the last row is
the same but with the cloud a factor of three longer in the LOS direction. All
calculations assume an external attenuating layer of $A_{\rm V}=1^{\rm mag}$.
}
\label{fig:plot_residuals_4}
\end{figure}

Instead of modifying the dust model or the radiation field, the temperature
gradients can be reduced by making the cloud more extended in the LOS
direction. Figure~\ref{fig:plot_residuals_4} shows alternative fits where, the
LOS density distribution corresponds $FWHM$ of 0.2\,pc, 0.5\,pc, or 0.9\,pc.
There is a clear difference between the $A_{\rm V}=0^{\rm mag}$ and $A_{\rm
V}=1^{\rm mag}$ cases, but the cloud $FWHM$ has an even stronger effect.  If
the cloud is made very elongated in the LOS direction with $FWHM$=0.9\,pc, the
fit to the 250-500\,$\mu$m data is quite good, although the 160\,$\mu$m
emission remains overestimated.

Although interstellar filaments are expected to be narrow with $FWHM\sim
0.1$\,pc, the $FWHM$ parameter describes the whole cloud, where values 0.2\,pc
and even 0.5\,pc might still be realistic. However, the use of Plummer
profiles that are derived directly from the fits to the POS filament structure
(column density) should result in a better simultaneous description of both
the narrow filament and the surrounding extended cloud.
Figure~\ref{fig:plot_residuals_4} shows that this default LOS Plummer profile
results in roughly similar quality of fit as the Gaussian model with
$FWHM=0.5$\,pc. By increasing the LOS extent up to an aspect ration of 3:1,
the residuals again drop to $\sim$4\% or below along the main filament.

\begin{figure*}
\includegraphics[width=18cm]{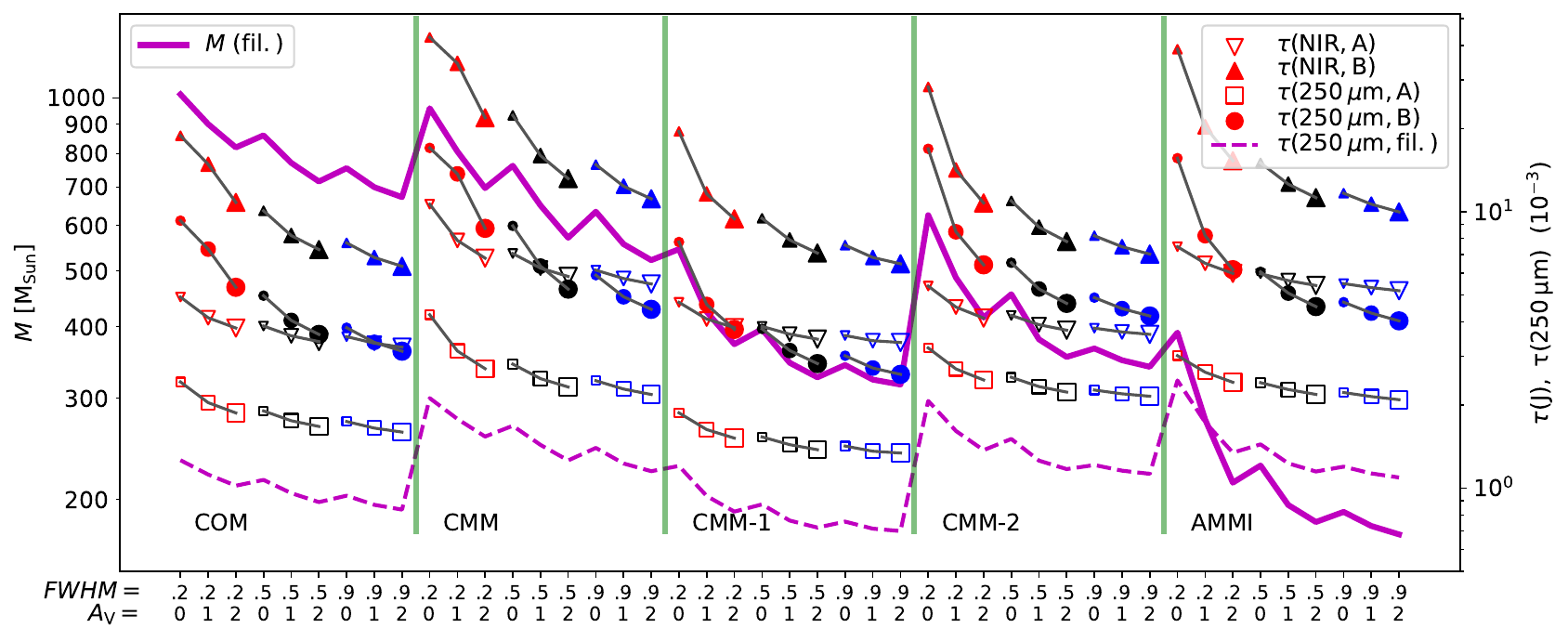}
\caption{
Mass and optical depth values in selected RT models. The solid magenta line
and the left axis show the estimated mass for the filament area, and the
dashed magenta line and the right axis the corresponding mean optical depth
$\tau(250\,\mu{\rm m})$. The symbols show the J-band (triangles) and
250\,$\mu$m (squares) optical depths (right axis) for the positions A (open
symbols) and B (filled symbols).
The red, black, and blue colours correspond to $FWHM$=0.2, 0.5, and 0.9\,pc,
respectively, and the small, medium, and large symbols to $A_{\rm V}$=0, 1,
and 2\,mag.
The x-axis is also labelled according to the $FWHM$ and $A_{\rm V}$ values of
the models.
}
\label{fig:plot_model_taus_3}
\end{figure*}

Although the quality of the fits can be similar, different assumptions (dust,
radiation field, cloud shape) result in significantly different predictions.
Figure~\ref{fig:plot_model_taus_3} compares the NIR and FIR optical depths and
the mass estimate of the fits that use different assumptions for the dust
properties, the $A_{\rm V}$ value of the attenuating layer, and the LOS extent
of the cloud (FWHM of Gaussian density distribution). The mass estimates
depend on the FIR emissivity but also indirectly on other factors that control
the dust temperatures. The estimates vary by a factor of five between the COM
(low FIR opacity) and AMMI (high FIR opacity) fits. However, this results
directly from the different absolute dust opacities (a factor of five between
AMMI and COM). The LOS cloud size is the second most important parameter, with
up to 50\% decrease in the mass estimates between the smallest $FWHM=$0.2\,pc
and the largest $FWHM$=0.9\,pc values. The effects of the radiation field
attenuation $A_{\rm V}$ are only slightly smaller and are particularly clear
for the more compact clouds (small $FWHM$). The mean optical depth
$\tau(250\,\mu{\rm m})$ varies by a factor of three and depends as much on the
cloud FWHM as the dust model. The NIR and FIR optical depths are naturally
correlated. There are thus similar large differences in the model-predicted
$\tau({\rm J})$ values, and direct NIR extinction measurements should be able
to rule out some dust models.

\begin{figure*}
\includegraphics[width=18cm]{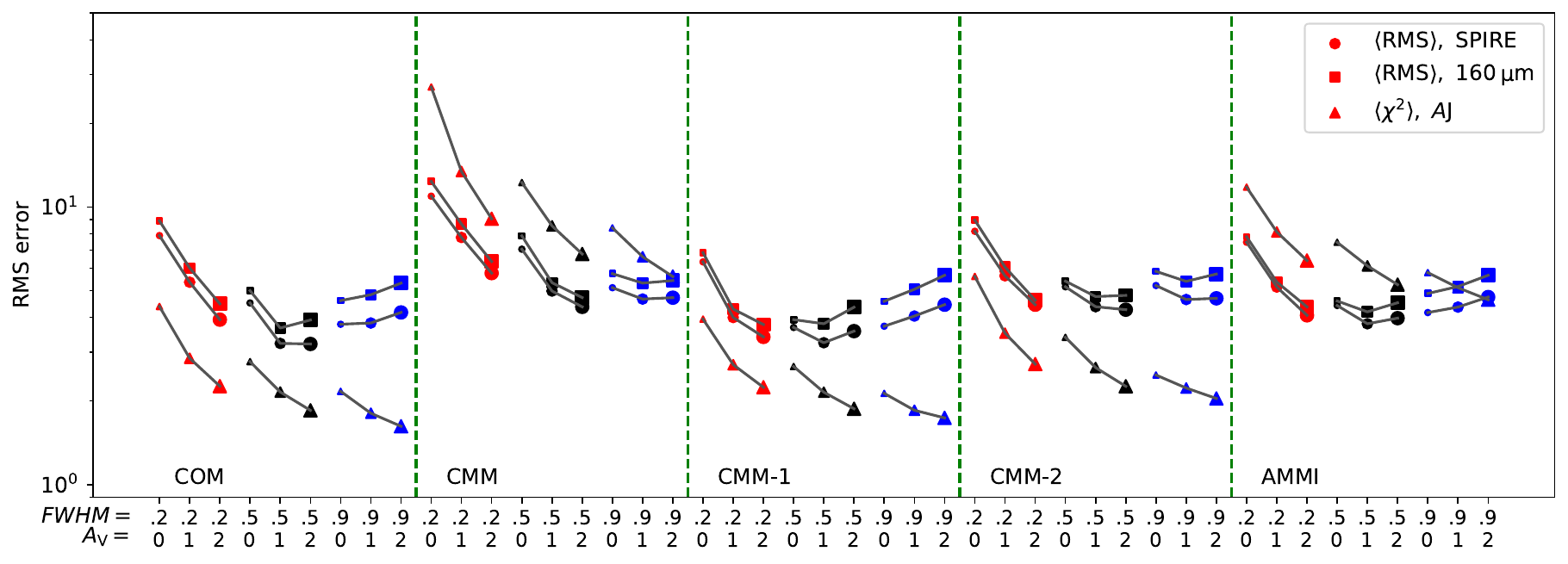}
\caption{
Errors in selected model fits with the COM, CMM, CMM-1, CMM-2, and AMMI dust
models. The errors are shown for the original SPIRE fit (circles) as well as
for the 160\,$\mu$m surface brightness (squares) and NIR extinction $A({\rm
J})$ (triangles). For the FIR data the values are the mean relative error over
the filament region. For the NIR extinction the value is the mean $\chi^2$
value that is calculated over the stars inside the filament area and using the
$A({\rm J})$ error estimates of the individual 2MASS stars. The red, black,
and blue colours correspond to $FWHM$=0.2, 0.5, and 0.9\,pc, respectively, and
the small, medium, and large symbol sizes correspond to $A_{\rm V}$=0, 1, and
2\,mag. The x-axis is labelled according to these parameters.
}
\label{fig:big_quality_1}
\end{figure*}

Figure~\ref{fig:big_quality_1} compares the fit quality for the models in
Fig.~\ref{fig:plot_model_taus_3}. This is shown as the mean rms value of the
relative error in the fitted 250-500\,$\mu$m bands, which is on average around
5\%. The plot additionally show the rms errors of the predictions of the
160\,$\mu$m surface brightness and of the J-band extinctions. The latter are
calculated using the individual 2MASS stars, normalised with the formal error
estimates of the $A_{\rm J}$ measurements. The SPIRE and 160\,$\mu$m errors
are correlated to such an extent that the addition of the 160\,$\mu$m does not
substantially change the conclusions reached with the SPIRE bands only. The
CMM models tend to result in the worst fit. The COM and CMM-1 models give the
best match to the SPIRE observations and the NIR extinctions. The CMM-2 and
AMMI models are slightly worse in matching the SPIRE observations, but,
compared to CMM, CMM-2 gives a much better match to NIR extinction data.

\begin{figure}
\includegraphics[width=8.8cm]{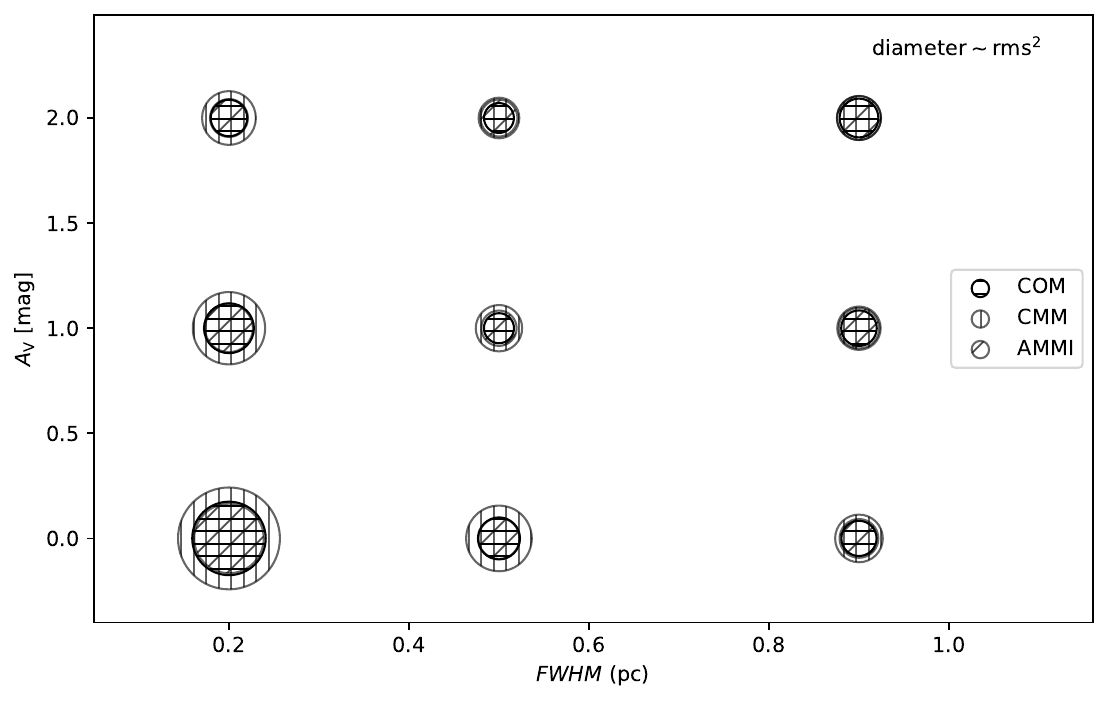}
\caption{
Relative rms error for the model fits using the COM, CMM, and AMMI dust models.
Errors are plotted against the $FWHM$ (x-axis) and $A_{\rm V})$ (y-axis)
parameters. The data correspond to region C (cf. Fig.~\ref{fig:plot_map}). The
diameter of the symbols is proportional to the squared error. Models with
larger LOS extent and/or more extincted radiation fields tend to result in
better fits.
}
\label{fig:plot_fit_quality_1}
\end{figure}

\begin{figure}
\includegraphics[width=8.8cm]{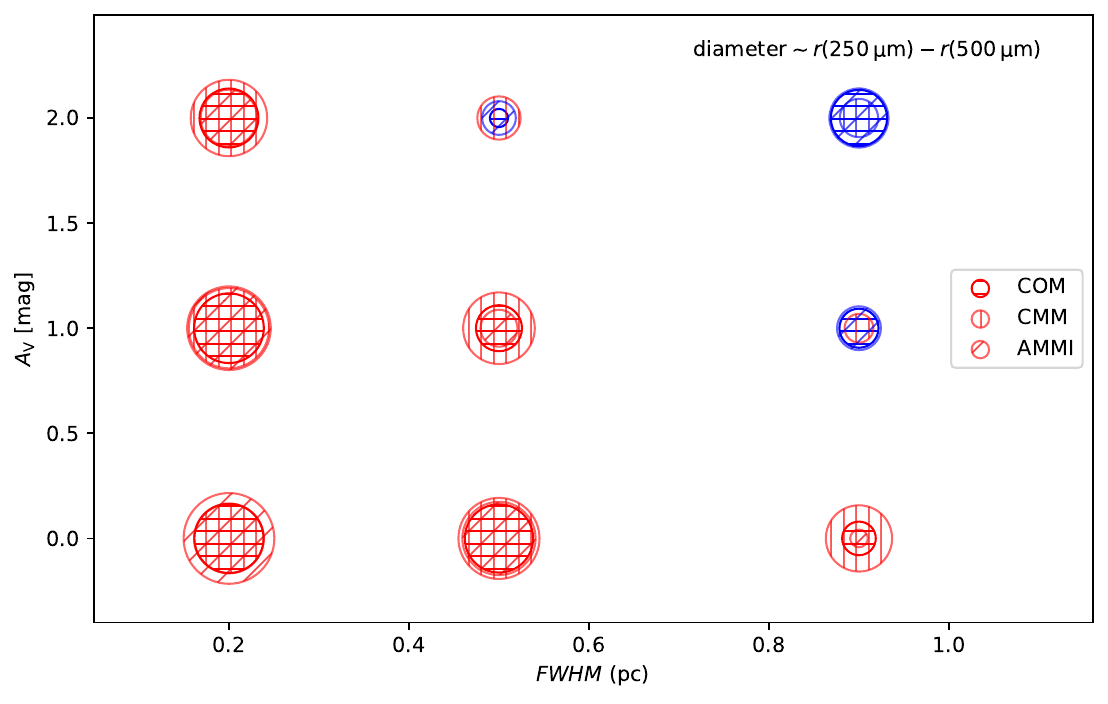}
\caption{
As Fig.~\ref{fig:plot_fit_quality_1} but showing the relative fit residuals
(250\,$\mu$m residual minus 500\,$\mu$m residual) in the region C. Red symbols
correspond to cases where the 250\,$\mu$m emission is underestimated and
500\,$\mu$m emission is underestimated towards the main core. In the case of
the blue symbols the situation is reversed: with larger LOS extent and large
external extinction, the models overestimate rather than underestimate the
dust temperature in the core.
}
\label{fig:plot_fit_quality_0}
\end{figure}

Systematic fit residuals would at first glance appear to be potential
indicators of dust evolution. However, the above results show that the results
are affected by multiple factors that are not related to the dust.
Figure~\ref{fig:plot_fit_quality_1} shows the fit quality in the region C as a
function of $FWHM$ and$A_{\rm V}$. Figure~\ref{fig:plot_fit_quality_0} is the
corresponding plot for the relative residuals $r$ (red and blue colours for
positive and negative mean values of $r(250\,\mu{\rm m})-r(500\,\mu{\rm m})$,
respectively). As the parameters $FWHM$ and $A_{\rm V}$ are varied, the mean
residual changes sign. This applies to all tested dust models and takes place
over range of the ($FWHM$, $A_{\rm V})$ parameter space where the fit quality
changes only little. Obviously, this makes it more difficult to interpret fits
in terms of dust evolution. In contrast, the NIR extinction varies
significantly between the dust models but is less sensitive to the $FWHM$ and
$A_{\rm}$ parameters (Fig.~\ref{fig:big_quality_1}).

\subsection{Effect of embedded point sources} \label{sect:L1506_stars}

\begin{figure}
\begin{center}
\includegraphics[width=8cm]{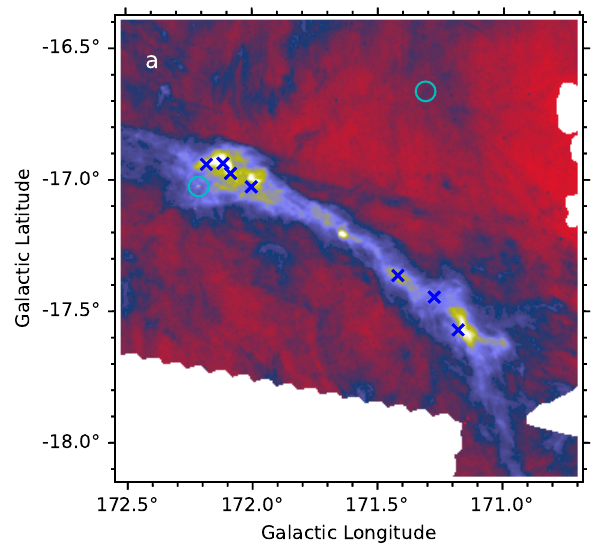}
\end{center}
\caption{
Locations of point sources plotted on the 250\,$\mu$m map of the L1506
filament. The cyan circles indicate positions of YSOs from \citet{Rebull2010},
and the blue crosses the locations of hypothetical embedded sources that were
added in the RT model.
}
\label{fig:plot_Rebull_A}
\end{figure}

The B212-B215 filament is not associated with bright stars, and YSO catalogues
lists only a couple of low-luminosity sources ($L<1 \, L_{\odot}$) that do not
coincide with the cores showing the largest fit residuals. However, internal
heating can significantly affect the SED of the protostellar cores relative to
the surrounding filament. We tested this possibility by adding a few ad hoc
point sources towards the intensity maxima in the filament of the Taurus model
(Fig.~\ref{fig:plot_Rebull_A}).

The blackbody temperature of the sources was set to either 200\,K or 3000\,K,
each run adopting the same value for all the sources. These temperatures were
used to mimic different types of sources, all of which remain unresolved in
our simulations. At the higher temperatures that emission is more towards the 
shorter wavelengths and, because of the higher optical depths, the effect
should be spatially more limited. The luminosity of each source was optimised
to result in zero average 250\,$\mu$m residuals within 30-70$\arcsec$ distance
of the source. The sources were places either at the exact location of the LOS
density maximum or displaced by 0.2\,pc along the LOS direction. The latter
should lead to the effect of the sources to be spatially more extended. Note
that the model cell size in dense regions is $\sim 0.006$\,pc, much below the 
0.2\,pc value.

\begin{figure*}
\begin{center}
\includegraphics[width=14cm]{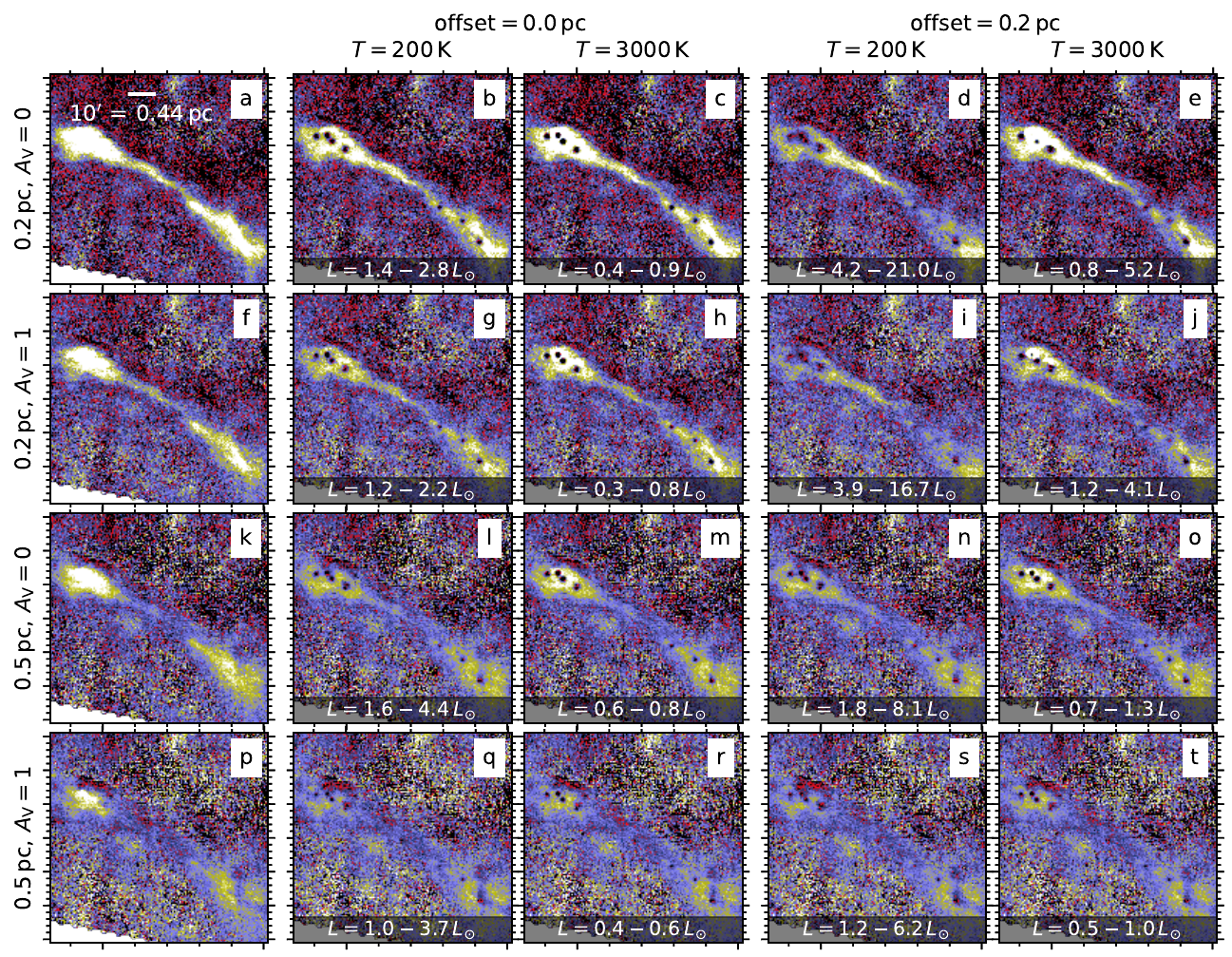}
\end{center}
\caption{
Comparison of 250\,$\mu$m fit residuals for models with hypothetical embedded
sources. Each row corresponds to a combination of $FWHM$ and $A_{\rm V}$
values, as noted on the left side of the frames. Each column of frames
corresponds to a different case concerning the point sources: first column
without sources, columns 2-3 with sources embedded at the location of the LOS
density maximum, and columns 3-4 with sources displaced 0.2\,pc along the LOS
from to the density maximum. The temperatures of the point sources are given
above each column of frames.
}
\label{fig:plot_PS_L1506}
\end{figure*}

Figure~\ref{fig:plot_PS_L1506} compares four models without point sources to
the corresponding models with point sources of different temperature and LOS
location. The residuals were previously seen to be largest for the smaller
$FWHM$ and $A_{\rm V}$. In these cases the effect of point sources remains
limited, partly because these models also tend to have higher optical depths
(at the wavelengths relevant for dust heating). Embedded sources are not able
to correct the extended positive residuals of the fits. At the same time,
their effect would be locally too strong, and the observations exclude the
possibility of the Taurus filament harbouring such sources
(Fig.~\ref{fig:plot_PS_L1506_b}). Only when the basic model already closely
matches the observations, such as with $FWHM=0.5$\,pc and $A_{\rm V}=1^{\rm
mag}$, embedded sources could provide minor improvement without being very
prominent in the frequency maps.

\subsection{Models with two dust components} \label{sect:L1506_twin}

\begin{figure}
\begin{center}
\includegraphics[width=8.8cm]{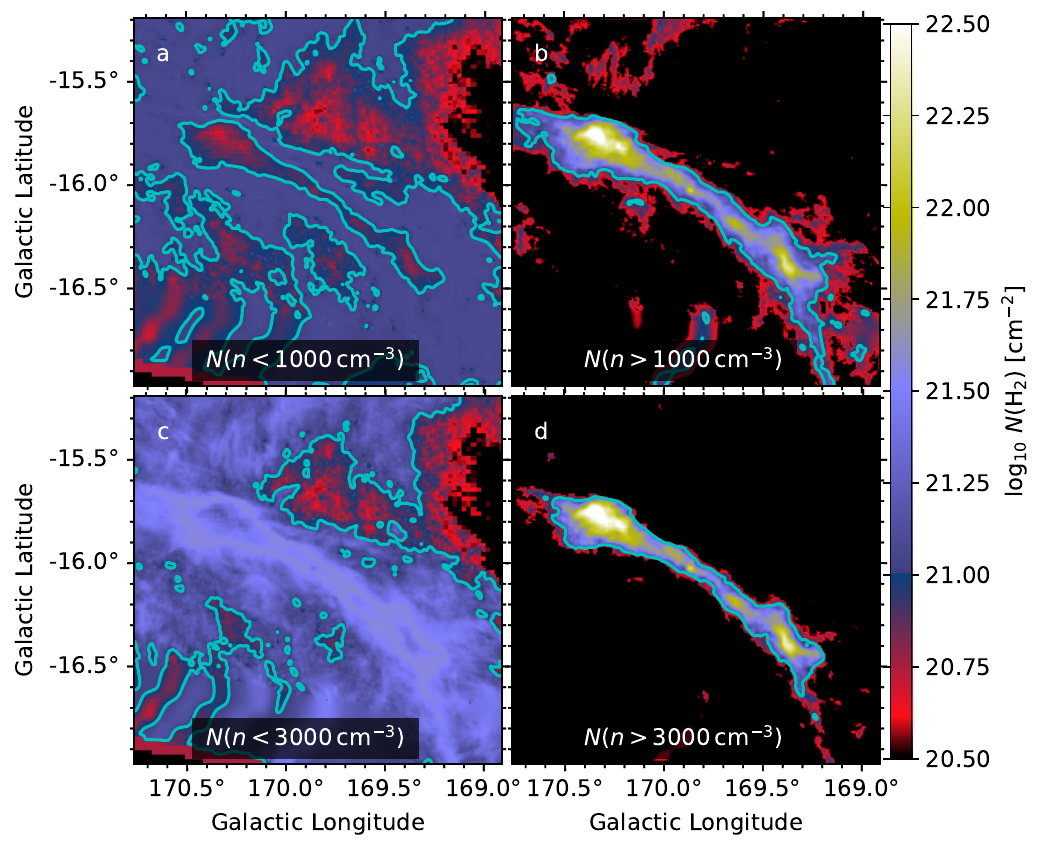}
\end{center}
\caption{
Examples of column densities associated with the dust components in two-dust
models. The plots correspond to a combination of CMM and AMMI dusts and
the model parameters $FWHM$=0.5\,pc and $A_{\rm V}=1^{\rm mag}$. The left frames
show the column density associated with the dust in regions of lower volume
density and the right frames the column density associated with the second
dust component. The density threshold for the transition between the
components is $n_{\rm 0}({\rm H})=1000\,{\rm cm}^{-3}$ in the upper and
$n_{\rm 0}({\rm H})=3000\,{\rm cm}^{-3}$ in the lower frame.  The cyan contour
is drawn at $N({\rm H})=10^{21}\,{\rm cm}^{-3}$.
}
\label{fig:plot_dust_colden}
\end{figure}

We examined next very briefly models with spatial variations in the dust
properties, using combinations of the COM, CMM, and AMMI dusts. We return in
Sect.~\ref{sect:discussion} to two-component models that include further
modifications to the dust properties. In the RT calculations, the dust
property variations are described as changes in the relative abundance of two
dust species (Eq.~(\ref{eq:abu})). The transition between the dust components
depends on the volume density and the selected density threshold $n_0$.
Figure~\ref{fig:plot_dust_colden} shows examples of how these translate to
spatial distribution of the components. Because the model densities are
adjusted during the model optimisation, the spatial distributions are slightly
different for each dust combination and $FWHM$ and $A_{\rm V}$ values.

\begin{figure}
\begin{center}
\includegraphics[width=8.8cm]{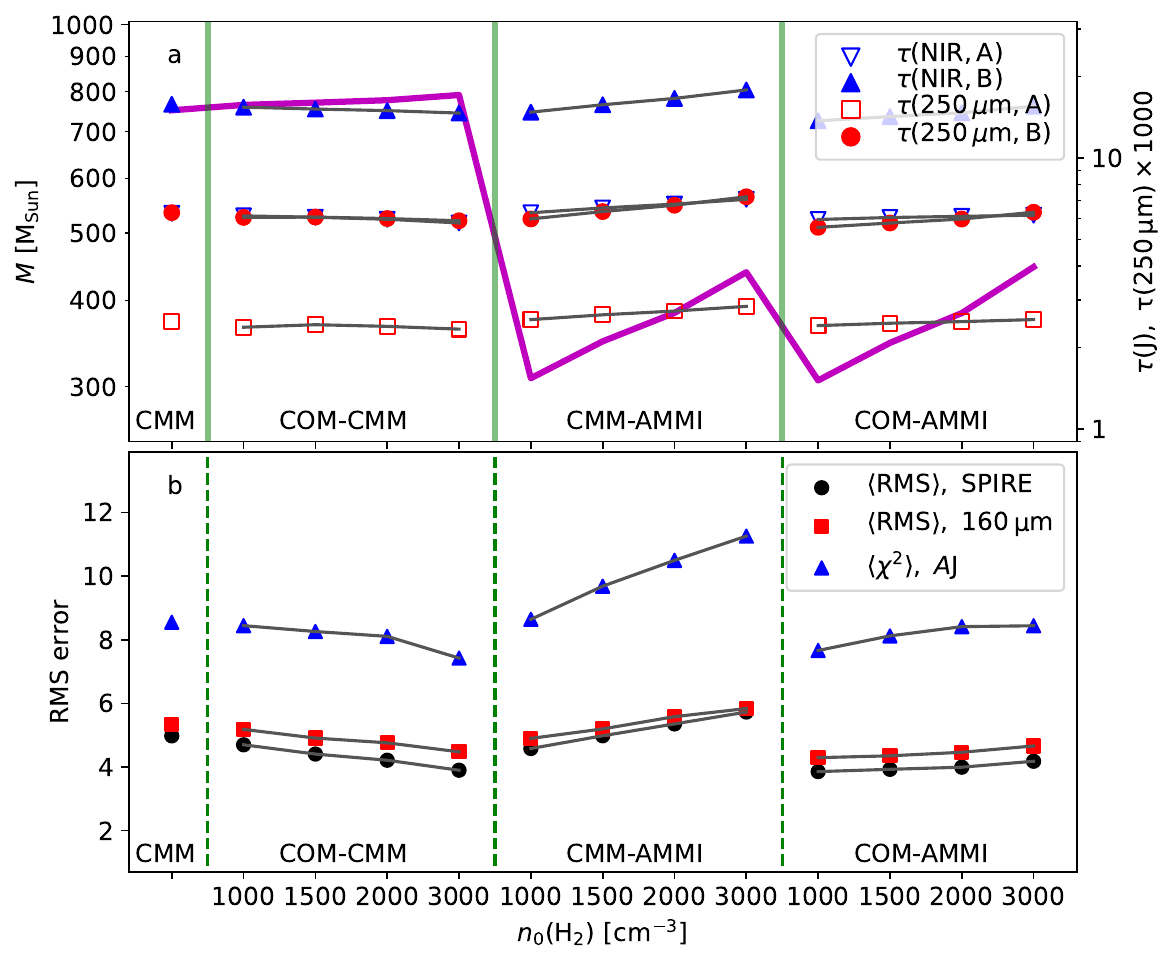}
\end{center}
\caption{
Comparison of mass and optical depth estimates (frame a) and the quality of
the fit (frame b) for models with two dust components. The results are plotted
for the single-component CMM model and three dust combinations (COM-CMM,
CMM-AMMI, and COM-AMMI), with density threshold $n_{\rm 0}({\rm H}_1)$ for the
transition between the two dust properties. Figure is similar to
Figs.~\ref{fig:plot_model_taus_3}-\ref{fig:big_quality_1}, with results shown
only for the case of $FWHM$=0.5\,pc and $A_{\rm V}=1^{\rm mag}$.
}
\label{fig:two_dust}
\end{figure}

Figure~\ref{fig:two_dust} compares some two-component fits to the calculations
with the single CMM dust, all with $FWHM=0.5\,{\rm pc}$ and $A_{\rm V}=1^{\rm
mag}$. The figure is similar to
Figs.~\ref{fig:plot_model_taus_3}-\ref{fig:big_quality_1}, showing the
estimated masses, NIR and FIR optical depths, and the match to observations.
The quality of the FIR fits (to SPIRE data and the extrapolation to
160\,$\mu$m) varies only little. In the single-component tests of
Fig.~\ref{fig:big_quality_1}, both COM and AMMI resulted in better fits than
CMM. The two-component fits are consistent with this trend: the smaller the
relative abundance of the CMM dust, the better the fit. However, compared to
the previous single-component models, these dust combinations provide little
improvement in the FIR fits and can even increase the disagreement with the
NIR extinction data.

\subsection{Comparison with analytical estimates} \label{sect:analytical}

The RT modelling can serve many purposes, one of which is the study of the
cloud column density structure. Column densities can be estimated also more
easily using MBB fits (Sect.~\ref{sect:MBB}), but the two estimates will not
be the same. An RT model gives a self-consistent description of the
temperature variations within the target but does not fit the observed
intensities everywhere very precisely. In contrast, MBB fits adopt a very
simple model for the source (e.g. possibly a single temperature) but usually
match the intensity observations better. MBB fits can use more complex
assumptions but quickly suffer from the degeneracy between the fitted
parameters \citep{Juvela2023_MBB}. 

\begin{figure}
\includegraphics[width=8.8cm]{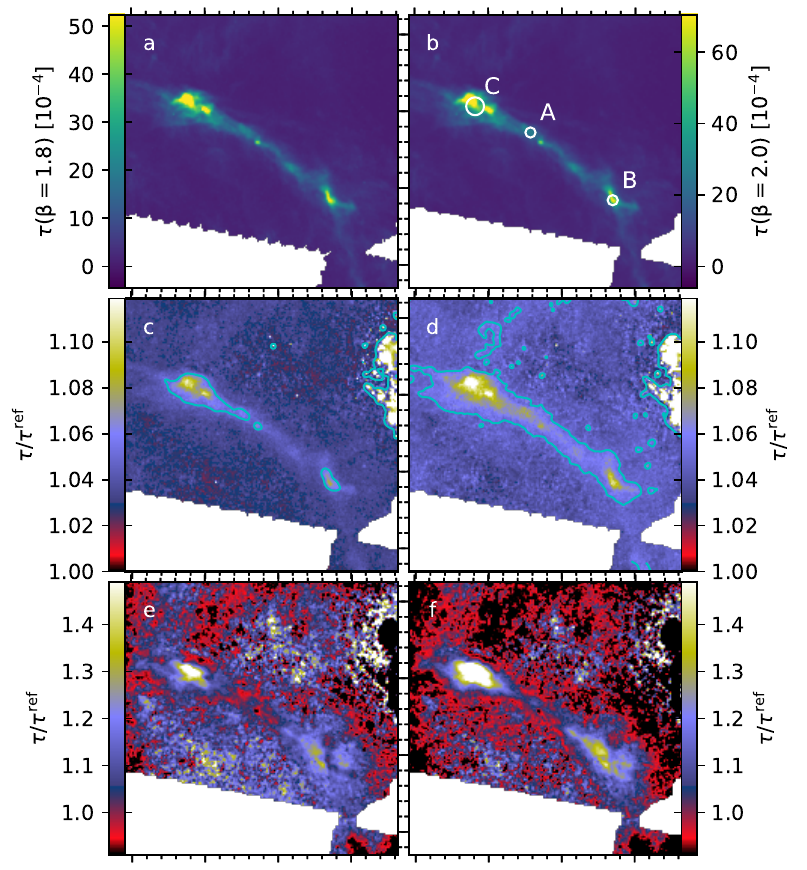}
\caption{
Comparison of 250\,$\mu$m optical depth estimates from MBB fits and RT models.
The uppermost frames show the results of single-temperature MBB fits with
$\beta=1.8$ (frame a) and $\beta=2.0$ (frame b). 
The second row shows the ratios between the MBB fit with a Gaussian
temperature distribution and the corresponding single-temperature fit from the
first row, with contours drawn at levels of $\pm$5\%.
The bottom row shows the $\tau$ ratios for two RT models ($FWHM=0.5$\,pc,
$A_{\rm V}=1^{\rm mag}$) and the single-temperature MBB fits from above. The
dust models are COM in frame e and CMM in frame f.
}
\label{fig:MBB_fits_1}
\end{figure}

Figure~\ref{fig:MBB_fits_1} compares optical depths with the SPIRE data
(250-500\,$\mu$m) at 41$\arcsec$ resolution. The estimates are from
single-temperature MBB fits, from MBB fits with a Gaussian temperature
distribution, and from two selected RT models. The single-temperature
calculations are taken here as the reference, although they are known to
underestimate the optical depths.

The MBB fits with Gaussian temperature distribution can still be done without
any priors, since they have only three free parameters (intensity, mean
temperature, width of the temperature distribution), the same as the number of
observed bands. As one piece of prior information, very low temperatures
$T_{\rm dust}<6.5$\,K were excluded from the proposed temperature
distributions. These fits result in higher $\tau$ values than the
single-temperature fits, although the difference are only about 10\% in the
densest regions.

Figure~\ref{fig:MBB_fits_1} includes two RT models, both with $FWHM=0.5$\,pc
and $A_{\rm V}=1^{\rm mag}$. These use COM and CMM dusts for which the $\beta$
values are, respectively, close to the $\beta=1.8$ and $\beta=2.0$ (the values
also adopted for the shown MBB fits). The COM model had given a relatively
good fit to the observations (Fig.~\ref{fig:big_quality_1}), and also in
Fig.~\ref{fig:MBB_fits_1} the estimates are outside the filament similar to
the two MBB fits. Along the filament, the values are close to the results of
the MBB fits Gaussian temperature distributions, although the map has more
noise-like fluctuations (relative to the MBB solution). The model optical
depths are higher towards the densest regions, by up to 30\% in the region B
and up to 60\% in the region C.

The CMM model provided a worse fit to the observations and was close to a
point where it could not produce the observed intensity levels, especially
near the C region, which can be expected to lead to higher $\tau$ values. The
predicted $\tau$ values along the filament are still similar to the MBB fit
with Gaussian temperature distribution solution, 5-10\% above the
single-temperature MBB fit. The values are higher towards the cores, by up to
40\% in around the region B and close to a factor of two near the region C. In
general, RT models should give more accurate (or more robust) estimates than
simple SED fits, but only if the models also accurately match the surface
brightness observations. In this case, the optical depths from the COM model
might be more accurate than the predictions of the CMM model, but this also
depends, for example, on the actual dust $\beta$ values in the cloud. The MBB
fits may look reliable, but the fact that they fit the observations to a high
precision does not mean that their optical depth predictions would be accurate
\citep{Juvela2023_MBB}.

\begin{figure}
\includegraphics[width=8.8cm]{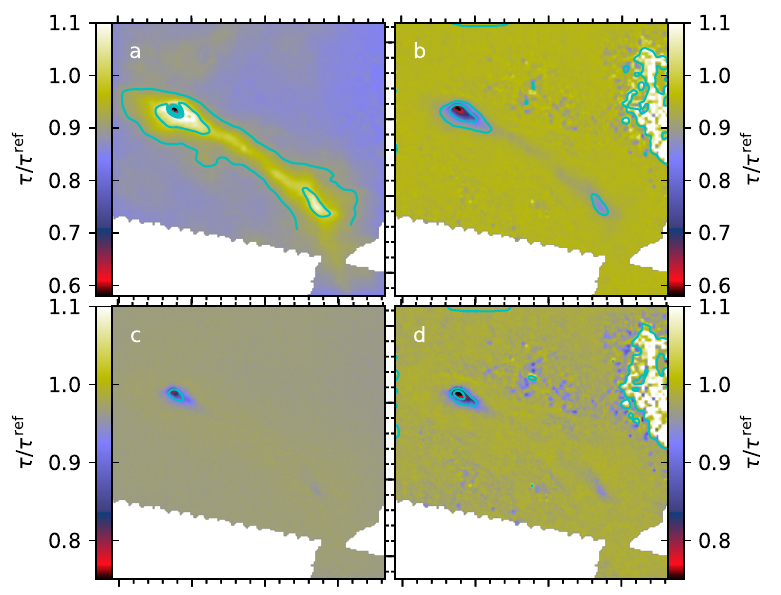}
\caption{
MBB optical depth estimates calculated for synthetic observations from RT
models. The uppermost frames show the results of MBB fits with a single
temperature component ($\beta=1.8$ in frame a, $\beta=2.0$ in frame b)
relative to the actual optical depths in the model. The second row shows the
corresponding ratios for MBB fits with Gaussian temperature distributions. The
RT models ($FWHM=0.5$\,pc, $A_{\rm V}=1^{\rm mag}$) used either the COM
(frames a and c) or the CMM dust (frames b and d). The cyan contours are drawn
at the values of 0.8, 0.9, and 1.0.}
\label{fig:MBB_fits_2}
\end{figure}

In the case of the Taurus observations above, the correct optical depths are
not known. Therefore it is useful to repeat the comparison in the other
direction, starting with the surface brightness maps predicted by the RT
models and comparing the MBB fit results to the known optical depths of the
models. The results are shown in Fig.~\ref{fig:MBB_fits_2}, as the ratio
between the MBB estimates and the now known true values. The results are
qualitatively similar to the analysis of Taurus observations above. The
single-temperature MBB fit underestimates the optical depth by up to
$\sim$40\%, and the errors of the MBB fit with Gaussian temperature
distributions is half of this and still in the same direction.

\section{Discussion} \label{sect:discussion}

The dust properties and dust evolution can be studied by comparing
observations to the predictions of RT models. In this paper, we have examined
this using the Taurus B212-B215 (LDN~1506) filament, to identify and quantify
some of the factors that may affect the reliability of the conclusions derived
from the modelling. 


\subsection{Model optimisation} 

In the RT modelling, we selected first the cloud shape, the dust properties,
and the SED shape of the radiation field (in terms of external extinction),
before optimising a set of other parameters. The optimisations were based on
heuristics, which is significantly faster than blind $\chi^2$ optimisation. 

The model column densities were adjusted based on the ratio of the observed
and model-predicted 350\,$\mu$m intensities, assuming that the surface
brightness is a monotonically increasing function of the column density.
Because column density is adjusted pixel by pixel, one could expect a perfect
fit at this wavelength, apart from small errors due to the model
discretisation, observational artefacts, or noise at scales below the beam
size. However, when the column density and radiation field were optimised, for
some dust models the intensities could be locally close to or even above what
the model could produce (e.g. Fig.~\ref{fig:plot_basic_fit}k). This depends
somewhat also on the surrounding area, because of the mutual shadowing of the
different parts of the cloud. Beyond the saturation point the surface
brightness decreases with increasing column density, and for this reason it is
important to start the model optimisation with low column densities combined
with high radiation field. The problem regions can be identified from positive
residuals at 350\,$\mu$m (and at longer wavelengths;
Fig.~\ref{fig:plot_residuals_4}c). 

If the radiation field were known, some models might be excluded already based
on this saturation and observations of a single frequency. However, we
included the strength of the external radiation field always as a free
parameter, $k({\rm ISRF})$. It was updated based on the average intensity
ratios $I_{\nu}(250\,\mu {\rm m})/I_{\nu}(500\,\mu{\rm m})$ in the filament
region, based on the knowledge that the ratio increases with increasing
strength of the radiation field. The optimisation results in the correct
average intensity ratio. This is similar but not exactly the same as the
maximum likelihood solution (minimisation of the $\chi^2$ value summed over
pixels). The distinction is not critical for the comparison of the models,
especially as the fit errors are dominated by systematic and spatially
correlated errors. Because the same scalar radiation field parameter applies
to the whole model and also the model density field is simpler than the real
cloud, a perfect fit to an extended cloud region is quite unlikely.

As described above, the optimisation always knows, based on the current
solution, in which direction the free parameters are to be adjusted.
Furthermore, although the number of column-density parameters was large (with
over 400 000 pixels per map), these are either independent (pixels located far
from each other) or strongly correlated (pixels within the same beam). This
results in fast convergence, and the calculations typically required only some
tens of optimisation steps. Since one RT run took about one minute or less
(down to 10\,s with a single dust component and without embedded point
sources), the model construction and optimisation is computationally quite
feasible. The model predictions were always convolved to the resolution of the
observations, which allowed one to used observational maps at different
resolutions. Basic single-temperature pixel-by-pixel MBB fits can be done for
small maps in a fraction of a second. However, if beam convolution is included
also into SED fitting, the run times increase and the computational advantage
relative to the full RT modelling may no longer be significant
\citep{Juvela2023_MBB}.

\subsection{Density fields}

Compared to the POS cloud shape, the LOS structure remains poorly constrained.
The LOS cloud size was also seen to have a significant effect on the dust
temperature variations. In the runs with constant dust properties, the more
extended models were clearly preferred for the Taurus filament. Fits to
250-500\,$\mu$m data required at least $FWHM$=0.5\,pc, but fits were the
better the larger the LOS cloud extent (Fig.~\ref{fig:big_quality_1}). Models
with Plummer density profiles gave similar results, calling for larger LOS
extent (e.g. with aspect ratio 3:1; Fig.~\ref{fig:plot_residuals_4}q-t).

Since the fit errors were largest in areas of high column density, also larger
$FWHM$ values are needed mostly in those areas. Figure~\ref{fig:plot_W} shows
examples of a further test, where $FWHM$ is optimised as a function of the map
position. A larger LOS size is indeed seen towards the cores, but the values
reaching $FWHM$=1.1\,pc, the maximum values allowed. In the case of the Taurus
filament, it is very improbable that the cloud would have such extreme
elongation and just in the LOS direction. This is not necessarily the case in
the analysis of isolated clumps, where the selection of bright but starless
sources might be biased towards elongated sources (even filaments) that are
seen along their major axis. Thus, in the absence of further information on
the volume densities, the unknown cloud structure can be a significant
hindrance on attempts to determine the dust properties.

In the case of the Taurus filament, the preference for longer LOS sizes could
in principle be related to the source inclination, since we have so far
assumed the filament to be in perpendicular to the LOS. There is some evidence
that the Taurus cloud is not aligned with the plane of the sky
\citep{Roccatagliata2020, Ivanova2021}, and for example \citet{Shimajiri2019}
suggested significant 20-70 degree inclinations for the neighbouring B211/B213
filament. Inclination can increase the LOS column density without a change in
the filament size perpendicular to its symmetry axis or any changes in the
dust temperatures. However, when the effect of inclination was tested in
practice, the fit quality changed only at $\sim$1\% level. Once the models are
optimised, the LOS optical depths remain quite similar and the lower optical
depth perpendicular to filament axis appear to be compensated by lower values
of the external radiation field (Fig.~\ref{fig:plot_inclined}). This is
similar to the conclusions of \citet{Ysard2013} regarding the effects of
inclination.

One final aspect of the cloud structure that was not examined in
Sect.~\ref{sect:results} is its potential small-scale inhomogeneity. This
would work in the same direction as a larger $FWHM$ and make the model at
large scales more isothermal. We tested this in the case of one model
($FWHM$=0.5\,pc, $A_{\rm V}=1^{\rm mag}$, CMM dust), multiplying the
originally smooth density field with a Gaussian random field (standard
deviation of 37\%, truncated at zero) that was generated for different
powerlaw power spectra, to compare the effects of mainly small-scale or
large-scale fluctuations. However, once the model was again optimised, the
differences to the original model were very small, both in terms of the fit
quality and the cloud mass and ISRF estimates. Only when the density
multipliers were squared (creating larger density fluctuations with a larger
fraction of low-density cells), the change became noticeable, but still only
at 10\% level of the original fit residuals. Thus, only extreme clumpiness of
the medium would alter the results.

Although the real LOS cloud extent cannot be directly measured, some limits
can be extracted from line observations. Based on the densities obtained by
modelling $^{13}$CO, C$^{18}$O, and N$_2$H$^{+}$ observations
\citep{Pagani2010_L1506}, \citet{Ysard2013} estimated for L1506C (i.e. around
the region B) an LOS width of at most $\sim$0.3\,pc. This is thus similar to
the filament POS thickness, and models with larger $FWHM$ (e.g. those with the
3:1 aspect ratios) appear in this case unlikely.

\subsection{Radiation field}

The radiation field was adjusted so to match the average SED shape in the
filament region. This does not prevent the fits from showing spatially
correlated errors at smaller spatial scales and at $\sim$10\% levels. The
250\,$\mu$m and 500\,$\mu$m errors are anticorrelated, as can be expected if
the dust is locally warmer or colder than in the real cloud. The residuals do
not show any obvious gradients over the field, which suggests that the assumed
ISRF anisotropy roughly matches the conditions in the cloud. Only the
160\,$\mu$m maps might show slightly larger positive residuals on the southern
side. If the ISRF were accurate outside the Taurus cloud, the part of the
molecular cloud that resides between the target region and the Galactic plane
could cause some shielding, which could contribute to this minor asymmetry.  

The errors were always mainly correlated with the column density. The tests
showed the spectral shape of the external radiation field has a clear effect
on the fits, although less than the cloud $FWHM$. Similar to larger LOS cloud
sizes, better fits were obtained by increasing the optical depth of the
external attenuating cloud layer. The mechanism is also the same: once the
short-wavelength photons are removed by the external layer, the model itself
is heated mainly at longer wavelengths where the optical depths are lower,
resulting in smaller temperature variations. The extinction was varied up to
$A_{\rm V}=3^{\rm mag}$ but, like the 3:1 aspect ratio for the cloud shape,
such high values are unlikely. In the \Planck 353\,GHz dust opacity map the
median value over the examined area is $3.7\times 10^{-5}$. 
To estimate a lower limit for the LOS extinction, the 353\,GHz value can be
first scaled to 250\,$\mu$m assuming $\beta=1.5$, further to J-band optical
depth using $\tau(250\mu{\rm m})/\tau(J)=1.6\times 10^{-3}$ \citep{GCC-V}, and
finally to visual extinction $A_{\rm V}=0.47^{\rm mag}$ assuming the $R_{\rm
V}=4.0$ extinction curve \citep{Cardelli1989}. 
On the other hand, the assumptions $\beta=2.0$, a factor of three lower
$\tau(250\mu{\rm m})/\tau(J)$ ratio (i.e. a value more consistent with diffuse
clouds), and $R_{\rm V}=3.1$ gives a value of 2.8$^{\rm mag}$. This applies to
the full LOS extinction and gives an upper limit $A_{\rm V}=1.4^{\rm mag}$ for
the extinction of the external layer. These are not strict limits either,
since they should describe the extinction towards the main radiation sources.
Due to the rest of the Taurus complex and other cloud, the extinction between
the target region and the Galactic plane could be higher that between the
target and the observer (or the Galactic centre).

\begin{figure}
\includegraphics[width=8.8cm]{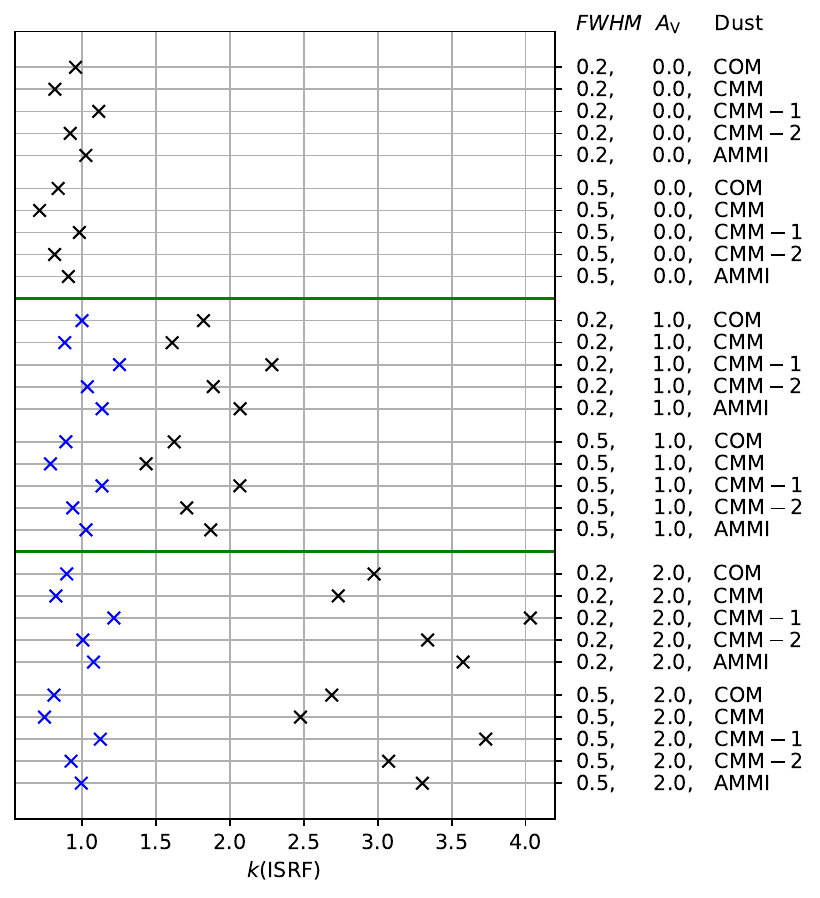}
%
\caption{
Radiation field scaling factors $k({\rm ISRF})$ in case of selected
single-dust models. The black crosses show the values for each combination of
cloud $FWHM$, $A_{\rm V}$, and dust model, as listed right of the frame. In
cases with $A_{\rm V}>0$, the blue crosses show approximate values at the
model boundary, if the effective attenuation is $e^{-\tau(0.8\,\mu{\rm m})}$,
using the 0.8\,$\mu$m optical depth of the external layer.
}
\label{fig:plot_ISRF}
\end{figure}

Each fitted model gives an estimate for the ISRF strength at the boundary of
the modelled volume or, for $A_{\rm V}>0^{\rm mag}$, outside the assumed
additional external layer. The estimates depend directly on the FIR spectral
index $\beta$ of the dust model and indirectly, via the dust energy balance,
on the optical/FIR dust opacities. Figure~\ref{fig:plot_ISRF} shows the ISRF
estimates for selected models. For $A_{\rm V}=0^{\rm mag}$ the values of
$k({\rm ISRF})$ are close to one, with a scatter of some 20\% and approximate
agreement with the \citet{Mathis1983} ISRF model. 

When $A_{\rm V}$ is assumed to be larger, the $k({\rm ISRF})$ values are
naturally larger since they describe the strength of the original field, not
the field at the model boundary. According to Fig.~\ref{fig:plot_ISRF}, in
these cases the $k({\rm ISRF})$ values are higher approximately by a factor
$e^{\tau}$, where $\tau$ is the optical depth of the external layer at
$\sim$0.8\,$\mu$m. The $0.8\,\mu{\rm m}$ optical depth thus seems to describe
the effective impact of the attenuation on the dust heating. The wavelength is
of course not constant but moves to larger values if the extinction is further
increased. If the external attenuation is $A_{\rm V}=1^{\rm mag}$, the field
outside the external layer is already twice as strong as the Mathis field.
Thus, in spite of the better match to the FIR observations, this makes models
with $A_{\rm V}>1^{\rm mag}$ clearly less likely, these also being in 
disagreement with the direct background extinction estimates derived from
\Planck observations.

\subsection{Errors in observations}

Before discussing potential spatial variations of dust properties, it is worth
noting that also systematic errors in the surface brightness measurements can
result in variations that are correlated with the observed intensities. These
could thus be misinterpreted as physical effects that would be connected to
the column density and volume density variations. This is especially true for
zero-point errors in the intensity measurements.

\begin{figure}
\includegraphics[width=9cm]{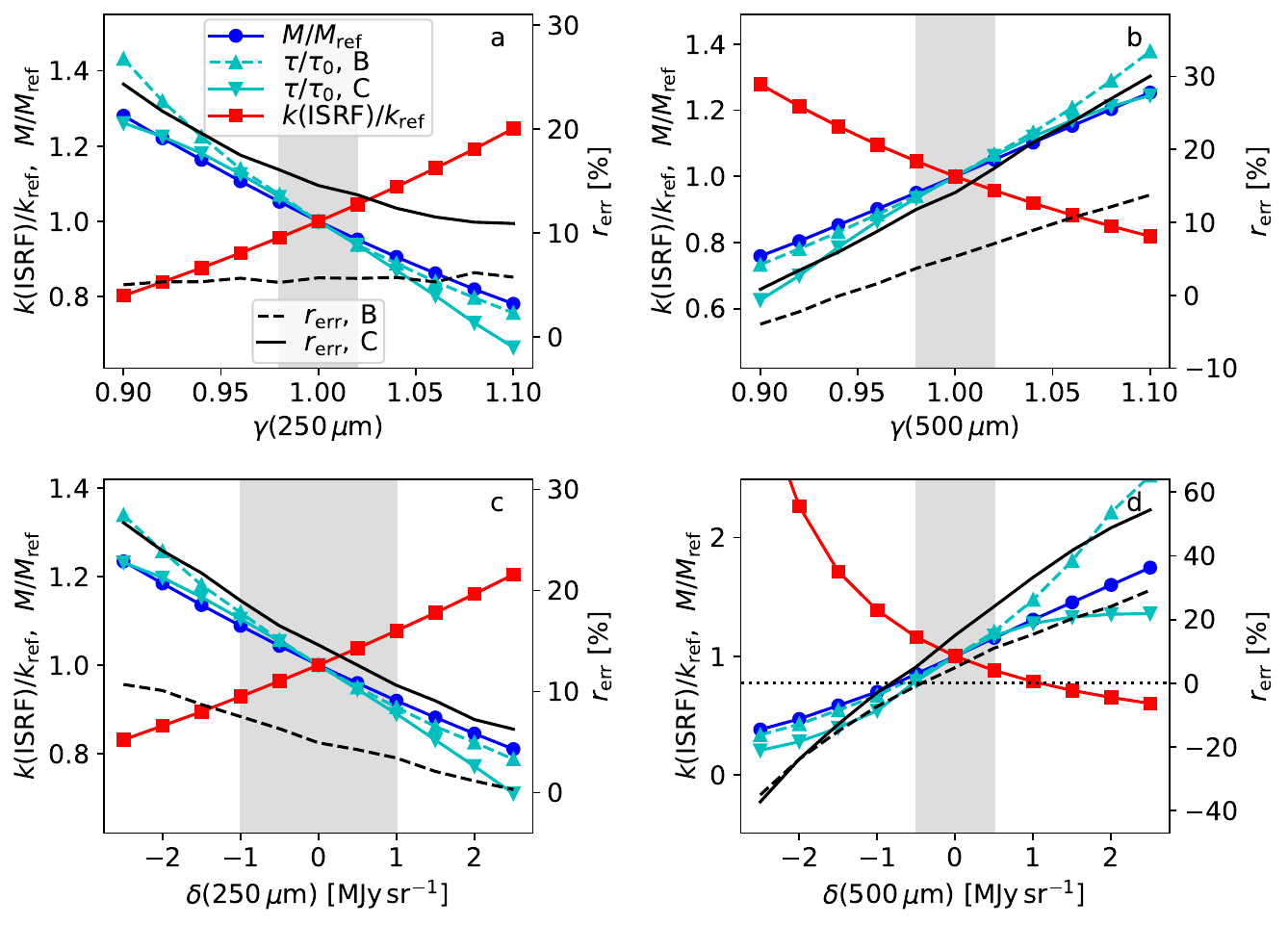}
%
\caption{
Effects of systematic observational errors. Errors are introduced to the
intensity measurements, and the frames show the effects on the filament mass
$M$, radiation field intensity, and optical depths $\tau$ in the sub-regions
B and C. The parameters are plotted relative to those estimated with the 
original observations. Each frame also shows the average 250\,$\mu$m residuals
in the B and C regions (right y axis, dashed and continuous solid lines).
The upper frames include multiplicative errors $\gamma$ in the 250\,$\mu$m
(frame a) or 500\,$\mu$m (frame b) measurements. The lower frames include
additive errors $\delta$ in the same bands. 
The shaded regions indicate probable upper limits limits for the errors in the
Taurus observations: 2\% for relative calibration and (ad hoc)
$\pm$1\,MJy\,sr$^{-1}$ and $\pm$0.5\,MJy\,sr$^{-1}$ for the 250\,$\mu$m and
500\,$\mu$m zero points, respectively.
}
\label{fig:plot_systematic_errors}
\end{figure}

Figure~\ref{fig:plot_systematic_errors} uses the model with $FWHM=0.5$\,pc,
$A_{\rm V}=1^{\rm mag}$, and CMM dust to illustrate the potential effects. An
error is assumed to affect either the scaling or the zero point of the
intensity measurements, either at 250\,$\mu$m or 500\,$\mu$m. The
multiplicative errors (factor $\gamma$) are varied between -10\% and 10\%, and
the additive errors in the range $\delta=\pm$3\,MJy\,sr${-1}$.  For the SPIRE
observations, the multiplicative errors should be below 2\%. Additive errors
could appear because of mapping artefacts (leading to errors in the difference
relative to the reference regions), statistical errors in the background value
derived using the reference region, or deviations from the assumption that the
background SED is constant over the field. In this paper, the background
subtraction was carried out using a relatively close reference region that has
an absolute 350\,$\mu$m surface brightness of at least 10\,MJy\,sr$^{-1}$
(Fig.~\ref{fig:plot_map}). For the sake of discussion, shading in
Fig.~\ref{fig:plot_systematic_errors} corresponds to ad-hoc upper limits of
1\,MJy\,sr${-1}$ at 250\,$\mu$m and 0.5\,MJy\,sr${-1}$ at 500\,$\mu$m for
these errors.

Compared to region B, the surface brightness in region C is much closer 
to the level where model predictions could start to saturate, and the fits fit
might behave there in an unexpected way. However,
Fig.\ref{fig:plot_systematic_errors} shows that the behaviour is both similar
and very systematic in both positions. Therefore, also the estimated mass of
the filament region is not unduly dependent on the fit in any single
sub-region. Any errors that make the SED to appear hotter naturally decrease
the mass estimates, but the effect of 2\% multiplicative errors is less than
10\% in the mass.

Additive errors are particularly insidious because they introduce
intensity-dependent changes in the band ratios, and they can even change the
sign of the residuals. In Fig.\ref{fig:plot_systematic_errors}, this is seen
only when the 500\,$\mu$m measurements have zero-point errors. When the error
is below $\delta=$-0.5\,MJy\,sr$^{-1}$, the residuals in both regions B and
C approach zero and they become negative for larger zero-point errors. Thus,
before fit residuals can be interpreted as a sign of dust evolution, one has
to be confident of sufficient absolute accuracy of the surface brightness
measurements. In the Taurus data, it is unlikely that zero-point errors would
explain all of the fit residuals (i.e. errors are likely to be well below
0.5\,MJy\,sr$^{-1}$ at 500\,$\mu$m). However, they could still have a
noticeable effect on the magnitude of the residuals, thus complicating the
interpretation of the observations in terms of dust property variations. More
secure conclusions could naturally be reached by investigating a set of
fields, provided that all observational errors average out in a large samples.

Some zero-point errors can potentially result also from spatial variations in
the SED shape of the background emission. This is not likely to be significant
factor in the Taurus field, due to its high Galactic latitude (little LOS
confusion) and the larger intensity contrast between the filament and the
reference region. However, for compact sources this could result even in
systematic effects, if the reference region is affected by limb brightening,
leading to overestimation of the short-wavelength emission in the reference
area compared to the target \citep{Menshchikov2016}.

\subsection{Dust properties}  \label{disc:dust}

Given the effects of all the other factors listed above, can something still
be said about the dust properties? 
In the basic single-dust models, the CMM model provided the worst fit to both
the FIR data and the NIR extinction, followed by the AMMI and the COM models.
The origin of the differences seems to be the same as for the other mechanisms
above, lower temperature contrasts leading to better fits. First, the spectral
index $\beta$ is lower for COM that the other two models ($\Delta \beta
\sim$0.2), which tends to lead to higher temperatures and thus lower optical
depths. Since the column densities were free parameters, the absolute level of
the dust opacity is not important but the balance between the NIR and FIR
opacities is. The ratios $\tau(250\,\mu{\rm m})/\tau({\rm J})$ are for CMM and
AMMI around $4.0 \times 10^{-4}$ and slightly higher $4.9 \times 10^{-4}$ for
COM (Fig.~\ref{fig:plot_dusts}). The COM models have therefore a lower optical
depth at the short wavelengths, also leads to slightly a better match to the
observed NIR values. The CMM-1 and CMM-2 modifications of the CMM model both
go in the same direction, where the $\Delta \beta=-0.3$ of CMM-1 had a larger
positive effect than the 50\% increase in the FIR opacity of CMM-2. The latter
correspond to $\tau(250\,\mu{\rm m})/\tau({\rm J})=6 \times 10^-4$, which is
still below the value of $\tau(250\,\mu{\rm m})/\tau({\rm J})=1.6 \times
10^{-3}$ suggested by some observations of dense clumps \citep{GCC-V}.

\begin{figure}
\includegraphics[width=9cm]{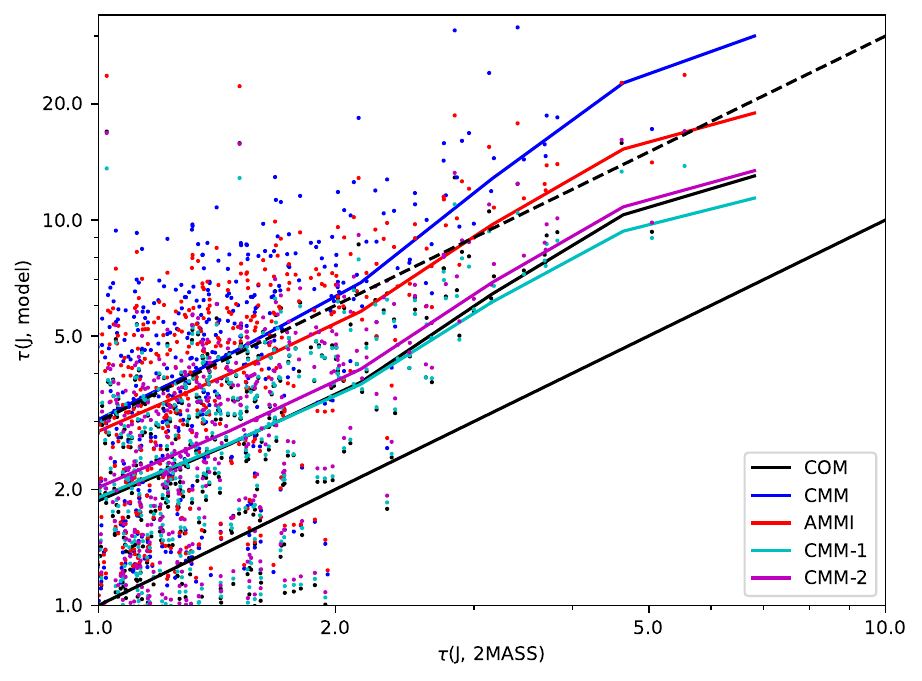}
\caption{
Model predictions for $\tau({\rm J})$ plotted against the estimates calculated
for 2MASS stars. The dots represent individual stars in the filament region
and the solid lines moving averages. The models correspond to $A_{\rm
V}=1^{\rm mag}$ and $FWHM=0.5$\,pc and the dust models listed in the legend.
The two modifications of CMM are CMM-1 ($\Delta \beta=-0.3$) and CMM-2 (50\%
increase in FIR opacity). The uncertainty of the relative zero points of the
data sets is expected to be a fraction of $\tau({\rm J})=1$, with little
impact especially in the high-$\tau$ end of the plot. The straight black solid
and dashed lines indicate, respectively, the one-to-one relationship and
values higher by a factor of three.
}
\label{fig:plot_NIR_quality}
\end{figure}

It is particularly noteworthy that all of the examined dust models 
overestimated the measured NIR opacities by a large margin. This is
illustrated in Fig.~\ref{fig:plot_NIR_quality}, where we plot the model values
against the $\tau({\rm J})$ estimates for the 2MASS stars. CMM is furthest and
the COM and CMM-1 models closest to the observed values. The error is larger
for CMM than for COM and AMMI, but the discrepancy is always significant, a
factor of 2-3. This does not directly imply a similar error in the dust
opacities, because the results in Fig.~\ref{fig:big_quality_1} depend in a
more complex way on the model optimisation (especially on the dust
temperatures).

If observations are to be fit with a single dust component, the spectral index
$\beta$ should be decreased or the ratio of the FIR and NIR opacities
increased. Direct measurements of $\beta$ are uncertain, but the typical
estimates are around $\beta=1.8$ for molecular clouds and possibly even higher
in FIR observations towards dense parts of molecular clouds
\citep{Sadavoy2013,GCC-VI,Bracco2017,Juvela_2018_pilot}. The spectral index of
the CMM-1 model already had a lower value of $\beta\sim 1.7$. Significantly
lower $\beta$ values are observed mainly at very small scales (e.g. in
protostellar sources, less relevant here) and at millimetre wavelengths
\citep{Planck2014_XVII,Sadavoy2016,Mason2020}. 
On the other hand, the dust opacity is believed to increase quite
systematically from diffuse medium to molecular clouds and further to cores
\citep{Martin2012,Roy2013,GCC-V}. In one of the earlier studies on Taurus,
\citep{Stepnik2003} concluded that the dust FIR/submm emissivity in LDN~1506
would be 3.4 times higher than in diffuse regions. Qualitatively similar
conclusions were reached even for the high-latitude cloud LDN~1642, which was
studied in \citet{Juvela_L1642}. There, in addition to FIR emission and NIR
extinction measurements, the need for dust with lower optical-NIR opacities
was also shown by the modelling of the optical-MIR scattered light.

\begin{figure*}
\begin{center}
\includegraphics[width=18cm]{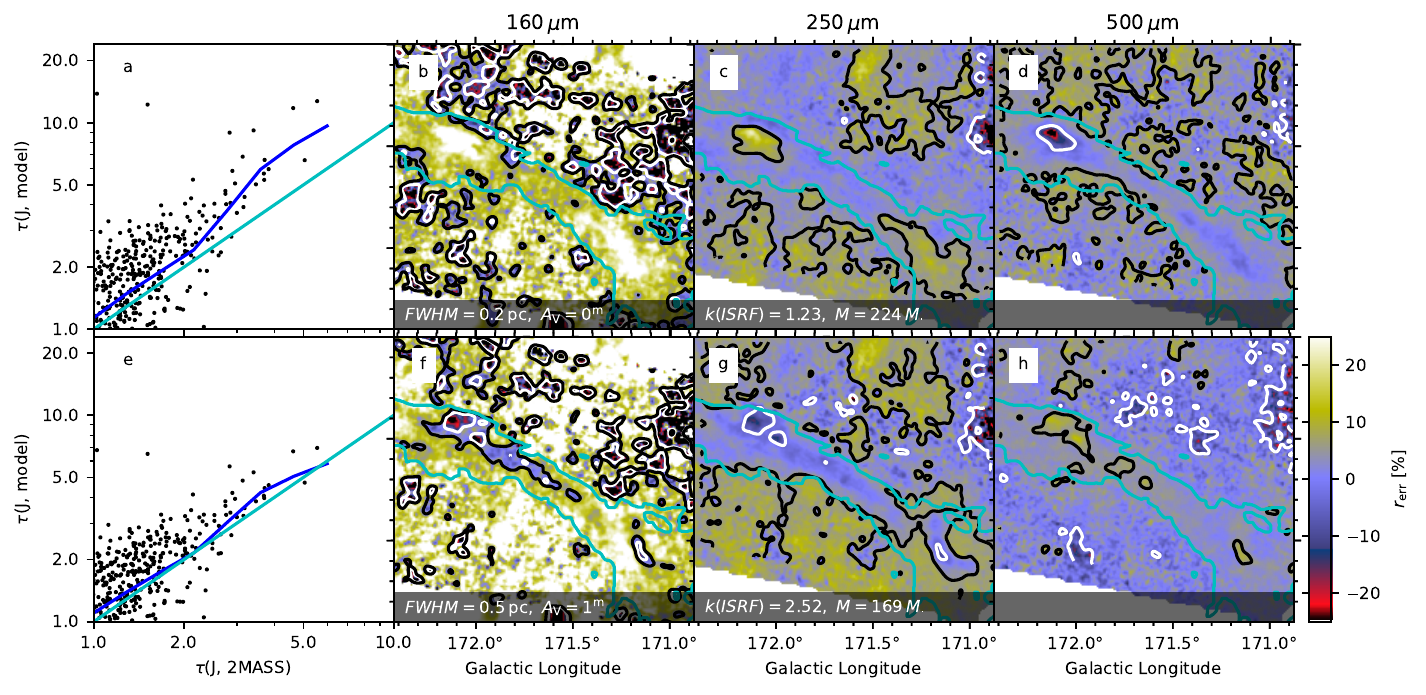}
\end{center}
\caption { 
Fits with the modified CMM-3 dust. The upper frames correspond to a
$FWHM=0.2$\,pc and $A_{\rm V}=0^{\rm mag}$ model and the lower frames to a
$FWHM=0.5$\,pc and $A_{\rm V}=1^{\rm mag}$ model. The leftmost frames show the
model-predicted J-band optical depths against the values measured with 2MASS
stars. Dots show the values for individual stars, the solid blue line is a
moving average, and the straight cyan line corresponds to the one-to-one
relation. The other frames show the 160, 250, and 500\,$\mu$m fit residuals
($r_{\rm err}$), where the -5\% and +5\% error levels are indicated with white
and black contours, respectively. It is noteworthy that in the regions B and C
(B212 and B215) the 250\,$\mu$m and 500\,$\mu$m residuals have different signs
in the two models.
} 
\label{fig:plot_CMM-3}
\end{figure*}

Based on the above, we tested a further modified dust model, CMM-3, where the
FIR $\beta$ was kept the same as in CMM but the FIR opacity was increased to
$\tau(250\,\mu{\rm m})/\tau({\rm J})=1.2 \times 10^{-3}$. This is a factor of
three higher than in the original CMM model but still below the observed value
quoted above \citep{GCC-V}.  Figure~\ref{fig:plot_CMM-3} shows that CMM-3
allows a better match to the FIR data, and the 250-500\,$\mu$m residuals
actually change signs between the case with $FWHM=0.2$\,pc and $A_{\rm
V}=0^{\rm mag}$ and the case with $FWHM=0.5$\,pc and $A_{\rm V}=1^{\rm mag}$.
This applies even at 160\,$\mu$m, and the predictions for the NIR optical
depth are much closer to the observations.

Thus, the observations can be explained with a model with almost the expected
cloud shape and radiation field, simply by adopting the higher
$\tau(250\,\mu{\rm m})/\tau({\rm J})$ ratio. The 250-500\,$\mu$m residuals are
small ($\sim$10\%) also outside the filament. In other words, the results do
not show strong evidence for dust evolution between regions of different
density. The 160\,$\mu$m residuals are still positive (up to $\sim$20\%) but
especially in Fig.~\ref{fig:plot_CMM-3}b this applies to the whole field.

\subsection{FIR evidence of dust evolution}

Section~\ref{disc:dust} showed that models can match the 250-500\,$\mu{\rm m}$
observations well, simply by assuming a high FIR-to-NIR opacity ratio
(Fig.~\ref{fig:plot_CMM-3}). Significant further fine-tuning is possible with
the $A_{\rm V}$ and $FWHM$ parameters. The evidence for spatial dust property
variations is not strong even at 160\,$\mu$m, especially considering the
potential effects of changes in the density field. Many studies have shown
that higher FIR-to-NIR opacity ratios are found only in dense clouds. The lower
ratios of diffuse clouds are also well constrained and correspondingly encoded
in dust models \citep{Draine2003ARAA,Compiegne2011,Jones2013,Hensley2023}. The
dust properties vary even in diffuse medium \citep[up to 50\% in the
FIR-to-visual opacity ratio;][]{Fanciullo2015}, and larger changes can take
place still at large scales, in the transition to molecular clouds
\citep{Roy2013,GCC-V,Hayashi2019}.
It would therefore be surprising if the dust properties remained constant over
the whole area, which was well over 10\,pc$^2$ in size. 

In Sect.~\ref{sect:L1506_twin}, none of the two-component models with the COM,
CMM, and AMMI dust properties resulted in significant improvement in the fits.
The best combination was COM-AMMI, because those components are also
individually best in fitting the FIR data. The match to NIR extinction
measurements was not improved compared to the single-dust models. Therefore, a
combination of different dust populations has little effect, if the dust
components themselves are far from correct. 

The column density was optimised for every LOS separately. This results, at
least at 350\,$\mu$m, to a better fit than if the densities were assumed to
follow a more rigid analytical prescription. This may be particularly
important when one investigates changes in the dust properties. Otherwise
small differences between the assumed (analytical) density profiles and the
real cloud structure could translate into artificial dust property variations
in the models.

\citet{Ysard2013} used cylindrical radiative transfer models to examine the
FIR observations of selected cross sections along LDN~1506 filament. With a 
different background subtraction, the modelling concentrated on the central
$\pm$0.2\,pc part of the filament. Each cross section was optimised
separately, with slightly different values of the dust and cloud parameters. A
good match to both NIR extinction and FIR emission required changes in the
dust properties. The models included a transition from standard diffuse-medium
dust at low densities to aggregates above a sharp density threshold of
$\sim$1000-6000\,cm$^{-3}$, with a factor of $\sim$2 increase in the
250\,$\mu$m opacity. Only the low spectral index value of the adopted
aggregates, $\beta \sim 1.3$, caused some problems in fitting the 500\,$\mu$m
observations. The FIR-to-NIR ratios of the aggregate models tested in that
paper \citep{Ossenkopf1994, Ormel2011} were in the range $\kappa(2.2\,\mu{\rm
m}/\kappa(250\,\mu{\rm m})$=208-379. This correspond to $\tau(250\,\mu{\rm
m})/\tau({\rm J} = 1\times 10^{-3} \,-\, 1.9\times 10^{-3}$, values that are 
similar or higher than in our ad-hoc CMM-3 case and in rough agreement with
the earlier modelling work of \citet{Stepnik2003}. At low densities
\citet{Ysard2013} adopted the dust properties from \citet{Compiegne2011}.
However, one can note that even the dust models developed for more diffuse
Milky Way regions have significant differences in their FIR-to-NIR optical
depth ratios \citep{Guillet2018,Hensley2023}. 

\begin{figure*}
\begin{center}
\includegraphics[width=18cm]{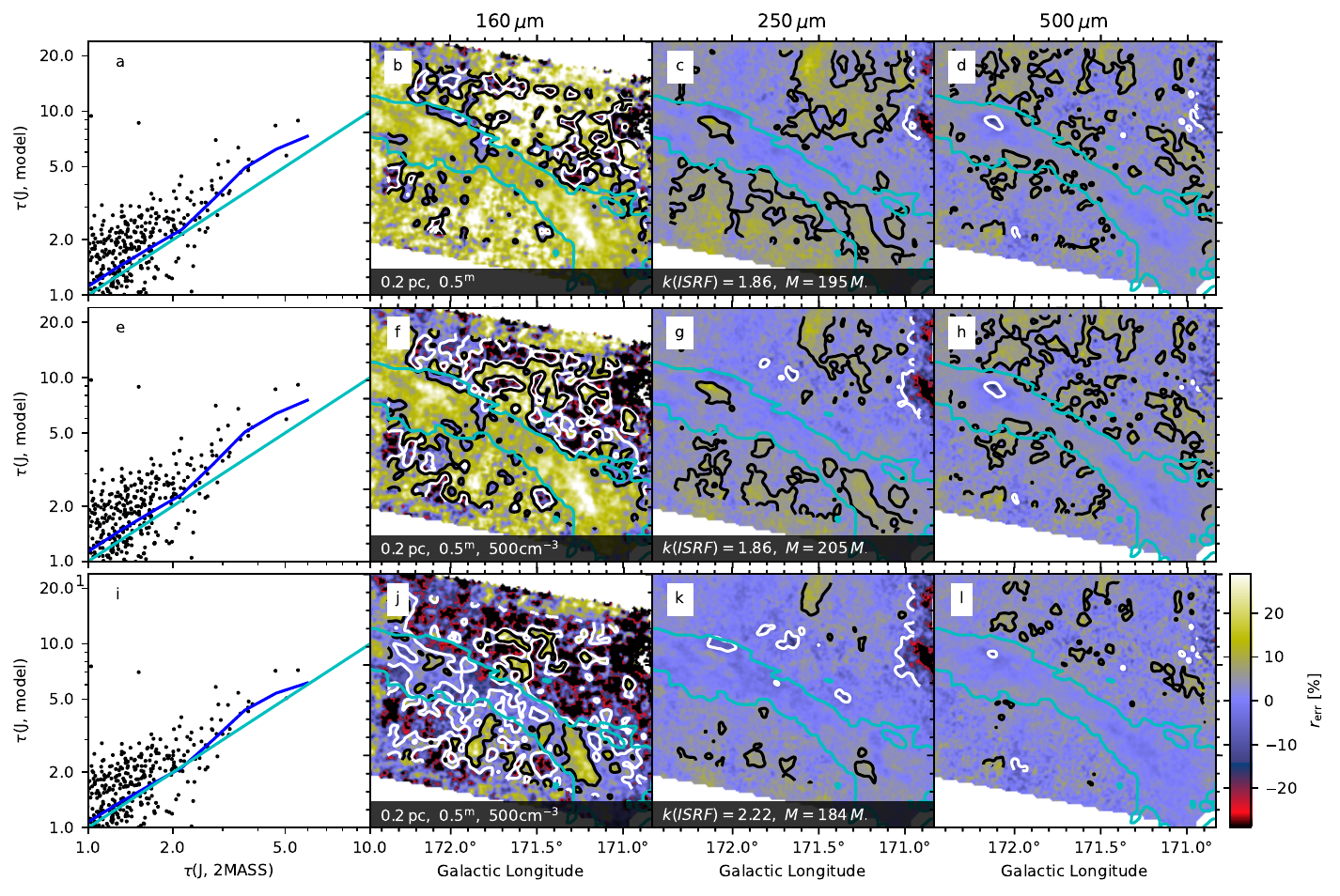}
\end{center}
\caption { 
Fits employing dust with large FIR emissivity. The leftmost frames plot the
model-predicted $\tau({\rm J})$ against the observed values. The other frames
show fit residuals at 160\,$\mu$m, 250\,$\mu$m, and 500\,$\mu$m, where the
-5\% and +5\% error levels are indicated with white and black contours,
respectively. The first row shows results for a model with CMM-3 dust, and 
the second row for a model using a combination of COM and CMM-3. The model on
the bottom row also uses two dust components, where CMM-3 is replaced with one
with a lower FIR opacity (twice the value in CMM) and a lower spectral index
($\beta\sim 1.7$). The model $FWHM$, $A_{\rm V}$, and $n_0$ values are listed
in the second and the $k({\rm ISRF})$ and mass estimates of the filament
region in the third frame column. 
}
\label{fig:plot_CMM-3_twin}
\end{figure*}

Our models try to find a self-consistent description for a larger cloud area.
Following the above examples, we also paired a diffuse-medium dust (COM) with
a dust variants with much higher FIR opacity. The other model parameters were
varied in the ranges $FWHM$=0.2-0.5\,pc, $A_{\rm V}$=0-1\,$^{\rm mag}$, and
$n_{0}$=0-3000\,cm$^{-3}$. For comparison, the first row of
Fig.~\ref{fig:plot_CMM-3_twin} shows a single-component fit with the CMM-3
dust, where the other model parameters are chosen to be between the two cases
of Fig.~\ref{fig:plot_CMM-3} and thus nearly optimal for that dust model. The
second row is a two-component fit that combines the COM and CMM-3 dusts. There
is marginal improvement at 250-500\,$\mu$m, and the 160\,$\mu$m residuals are
closer to zero, but only outside the filament. The models preferred a low
value of $n_{0}$, at there is little difference to the single-component fit.

We already saw in connection with single-dust models that both higher FIR
opacity and lower $\beta$ improved the fits. On the last row of
Fig.~\ref{fig:plot_CMM-3_twin}, COM dust is paired with another modification
of CMM. The FIR opacity is increased only by a factor of two (instead of the
factor of three in CMM-3) but the 250-500\,$\mu$m $\beta$ is also decreased to
1.72. The other parameters ($FWHM$, $A_{\rm V}$, $n_0$) are the same as in the
previous cases. The resulting fit to 250-500\,$\mu$m is better, although the
average errors were already below 5\%. The main change is at 160\,$\mu$m,
where the average residuals are closer to zero. The lower $\beta$ is
compensated by higher dust temperature, which has increased the 160\,$\mu$m
emission from the model and reduced the average residuals closer to zero.  

All three models in Fig.~\ref{fig:plot_CMM-3} are in good agreement with the
observations of the NIR extinction. After the adoption of the CMM-3 model, the
fits also prefer more compact clouds shapes ($FWHM<$0.5\,pc) and lower cloud
shielding ($A_{\rm V}\la 1^{\rm mag}$). These are both consistent with other
constraints from the measurements of the volume density (this also
corresponding to an approximate cylinder symmetry for the filament) and LOS
extinction. 

The high FIR-to-NIR opacity ratio makes the dust colder, which is compensated
in the models by a stronger radiation field.  These result in mass estimates
that are towards the lower end of all the single-dust models in
Fig.~\ref{fig:plot_model_taus_3}. All models in Fig.~\ref{fig:plot_CMM-3} have
$k_{\rm ISRF}$ values close to two, the parameter describing the field outside the
assumed external cloud layer. The estimated field strength also increases if
$\beta$ is decreased. In Fig.~\ref{fig:plot_CMM-3}, we chose to show the model
results for $A_{\rm V}=0.5^{\rm mag}$, which is in best agreement with the
estimates of the LOS extinction. The $A_{\rm V}=1^{\rm mag}$ models resulted
in even slightly better fits to the 160\,$\mu$m data. However, this increase
in $A_{\rm V}$ would also result in $k({\rm ISRF})$ increasing further by
0.8-1.0 units. The difference relative to models of the local ISRF may be
something of a problem already for the models in Fig.~\ref{fig:plot_CMM-3}.
Also in the L1506 models of \citet{Ysard2013}, attenuated fields resulted in
worse fits than the standard ISRF. However, in that case higher radiation
field values were not tested. The ISRF in the solar neighbourhood is well
constrained by direct and indirect observations and modelling
\citep{Mathis1983,Lehtinen2013,Fanciullo2015,PlanckXXIX2016,Mattila2018}. It
is still noteworthy that the DIRBE-observed average sky brightness is in NIR
some 50\% above the \citet{Mathis1983} values \citep{Lehtinen1996} and the
difference approaches at 5\,$\mu$m a factor of two. NIR radiation is
important for the dust heating in the deeper cloud layers. Nevertheless, since
the Taurus filament is not likely to get significant additional heating from
local radiation sources, a high $k({\rm ISRF})$ values of the models remain a
concern.

The ratio of FIR and NIR opacities is a sensitive tracer of dust
evolution. Apart from, for example, detailed spectroscopic observations of the
NIR-MIR extinction curve, also dust scattering across the same wavelengths can
provide useful constraints on the dust models
\citep{Lefevre2014,Saajasto2021,Juvela_L1642}.

\section{Conclusions} \label{sect:conclusions}

We have modelled the FIR dust emission over an extended region of the
B212-B215 filament in the Taurus molecular cloud. The goal has been to examine
how the modelling results are affected by different factors and how well the
dust properties can be constrained using FIR data over a limited wavelength
range. We also wanted to see, if it is possible to build a self-consistent
model for the larger, 16 square degree area with a large range of column
densities. The fits used primarily 250-500\,$\mu$m data, but the model
predictions for 160\,$\mu$m surface brightness and NIR extinction were also
examined. We used three basic dust models from the literature (COM, CMM, AMMI)
but finally also examined what further modifications of dust opacity and
opacity spectral index are needed in order to match the Taurus observations.

\begin{itemize}
\item 
The use of RT modelling for the analysis of extended maps has become feasible.
The three square degree maps were modelled using only $\sim$15 million cells,
with the run times of the individual RT runs varying between 10 seconds and
a couple of minutes.
\item
The models were optimised using simple heuristics for the column density and
radiation field updates. This allowed fast convergence and the final result
was usually found after some tens of iterations.
\item 
The use of the basic dust models led to small but significant differences in
the FIR fit quality. Largest errors are found observed towards regions of high
column density. However, the largest discrepancy was in the NIR extinction,
where these models overestimated the observed NIR extinction by a factor of
2-3.
\item 
The fits are strongly affected by the assumed LOS cloud size, especially in
regions of high column density, and by the spectral shape of the illuminating
radiation field. These effects are of similar magnitude or even larger than
the differences between the tested basic dust models.
\item 
The adoption of a different dust model may cause only small changes in the fit
quality and yet result in large differences in the estimates of the cloud mass
(up to a factor of five for the studied models) and the radiation field
intensity (up to $\pm$30\%). The mass differences are affected by the absolute
values of the dust opacity, while the estimates for the optical depths varied
by less than a factor of two.
\item 
The Taurus observations could be fit, down to the 160\,$\mu$m band, by
adopting a dust model where the FIR opacity was increased by a factor of 2-3
($\tau(250\,\mu{\rm m})/\tau({\rm J})=(0.8-1.2) \times 10^{-3}$).  The
resulting models also are in good agreement with NIR extinction observations,
but require a radiation field that is twice the standard value in the solar
neighbourhood.
\item
The models are consistent with the expected dust evolution, where for example
the formation of ice mantles and grain aggregates could explain the large FIR
dust opacities. However, although dust can be seen to be clearly different
from the normal dust in diffuse medium, the 250-500\,$\mu$m or even the
160-500\,$\mu$m data alone are not sufficient to unambiguously show dust
property variations within the field. 
\item 
The effects of dust property variations are partly degenerate with those of
the poorly constrained density and radiation fields. Systematic errors in
data, mainly in the zero points of the surface brightness measurements, can
produce further effects that are correlated with the column density. In the
case of Herschel data, the magnitude of these potential effects is small but 
not negligible.
\item 
The optical depths of the fitted radiative transfer models were compared to 
estimates from SED fitting. For the Taurus observations, single-temperature
MBB fits gave expectedly the lowest values, MBB fits assuming a Gaussian
temperature distribution up to 15\% higher values, and radiative transfer
modelling up to 50\% higher values. Similar results were observed when
analysing synthetic surface brightness observations from the models. However,
depending on the assumed dust properties, RT models may also overestimate the 
optical depths and the effect can be locally significant.
\end{itemize}

To constrain the dust models further, it would be important to have
multi-frequency data also on the NIR-MIR extinction and light scattering. Such
observations are currently possible with the James Webb Space Telescope
(JWST). As a higher mass counterpart to the Taurus filaments, the Orion
molecular cloud three (OMC-3) would be a promising target, due to the clear
MIR absorption seen in \Spitzer data towards its filaments
\citep{Juvela2023_OMC3}.

\begin{acknowledgements}

MJ acknowledges the support of the Academy of Finland Grant No. 348342.

\end{acknowledgements}

\bibliography{my.bib}

\begin{appendix}

\section{Additional plots} \label{app:p}

\begin{figure*}
\begin{center}
\includegraphics[width=12cm]{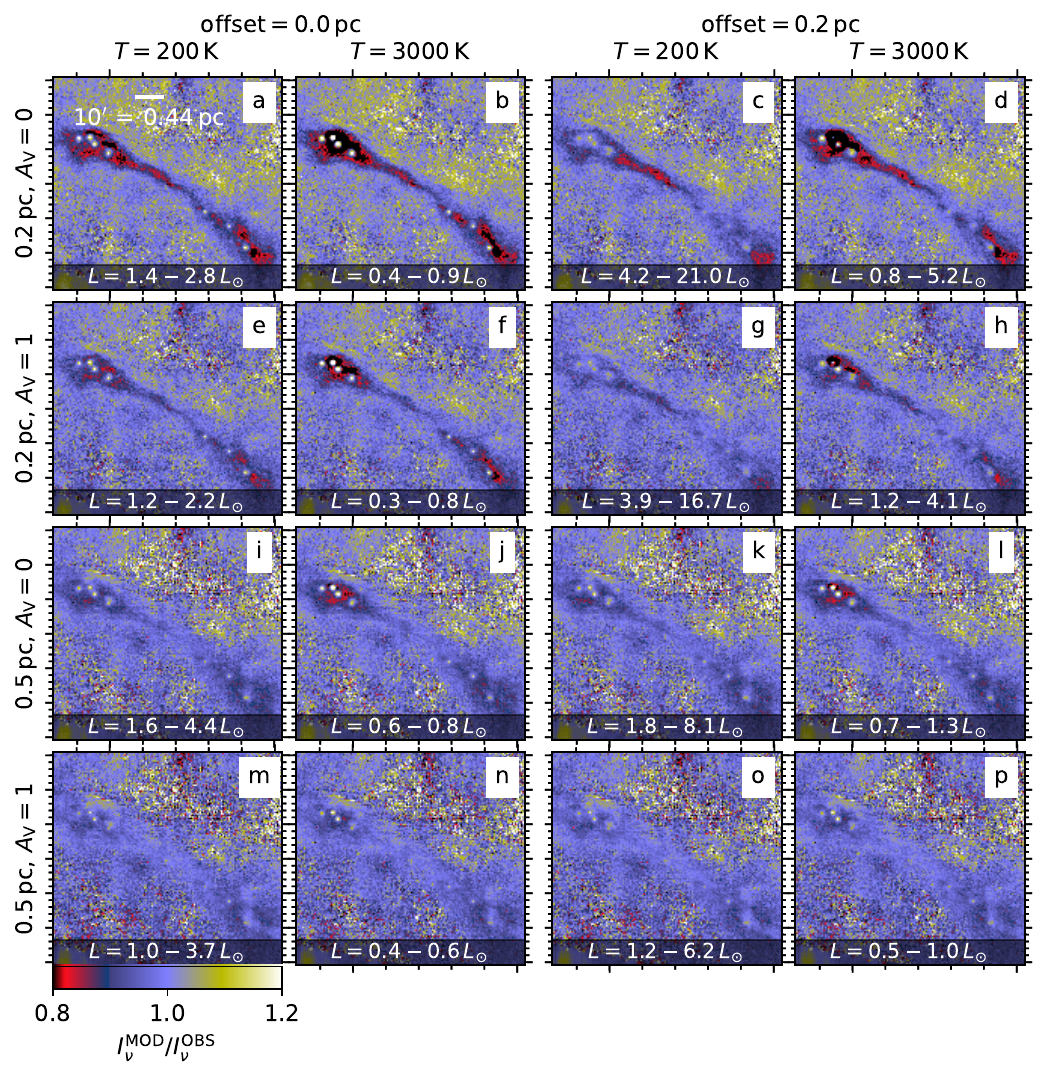}
\end{center}
%
\caption{
Ratio of modelled and observed 250\,$\mu$m surface brightness maps in the case
of models with ad hoc embedded sources. The models are the same as in
Fig.~\ref{fig:plot_PS_L1506}.
}
\label{fig:plot_PS_L1506_b}
\end{figure*}

Connected with the discussion in Sect.~\ref{sect:L1506_stars},
Fig.~\ref{fig:plot_PS_L1506_b} shows the ratio of the 250\,$\mu$m surface
brightness maps for models with embedded radiation sources that are divided by
the observed 250\,$\mu$m maps. The frames corresponds to the same models as in
the residual plot Fig.~\ref{fig:plot_PS_L1506}.  The presence of these
hypothetical embedded sources would be readily visible in the 250\,$\mu$m
data, and they would even more prominent at shorter wavelengths.

\begin{figure*}
\begin{center}
\includegraphics[width=14cm]{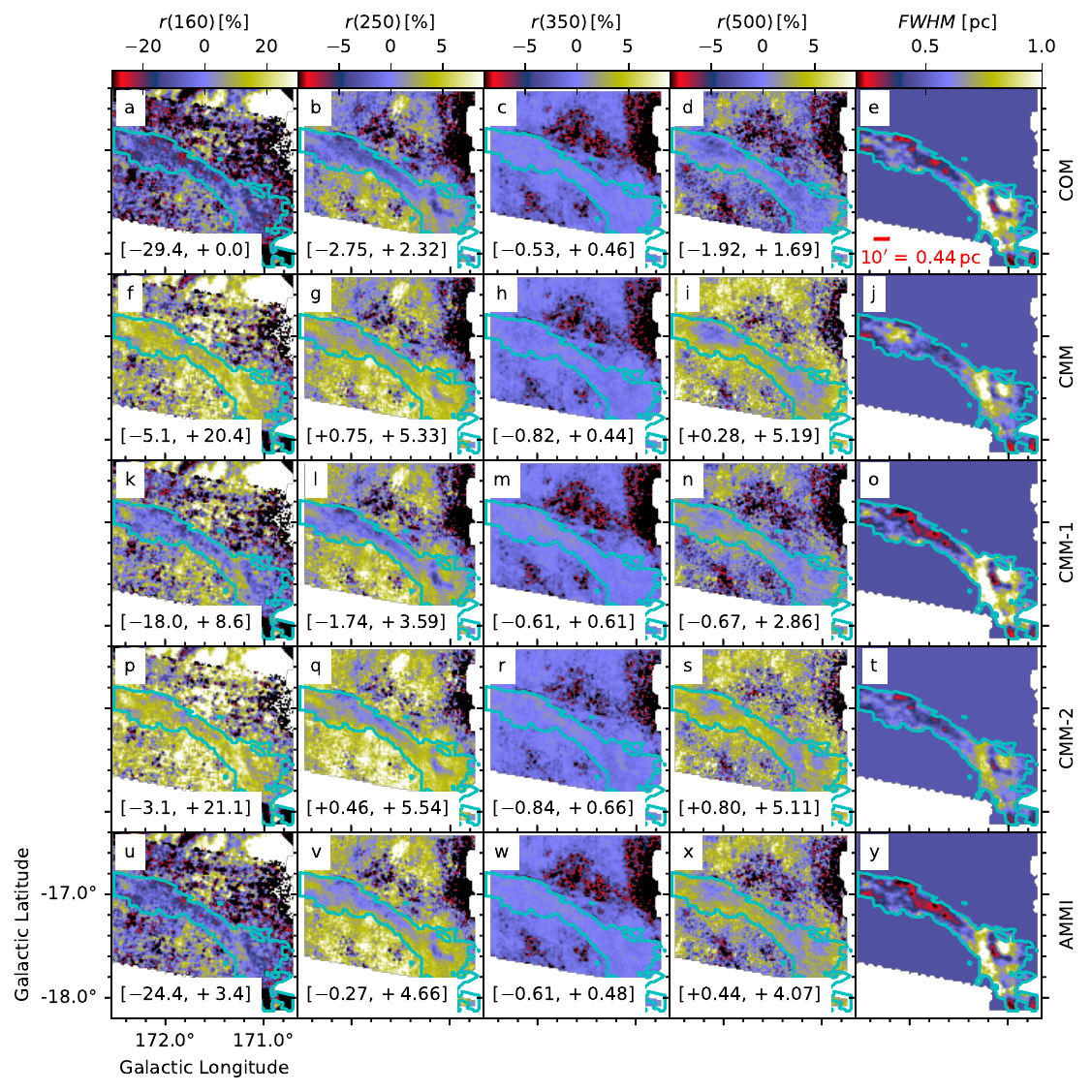}
\end{center}
%
\caption{
Examples of model fits where the LOS FWHM of the cloud is included as a free
parameter. On each row, the first four frames show the 160-500\,$\mu$m fit
residuals, and the final frame on the right the map of the FWHM scaling
(filament region only). The five rows correspond to the dust models COM, CMM,
CMM-1, CMM-2, and AMMI, respectively. All calculations were done assuming
$A_{\rm V}=1^{\rm mag}$. The cyan contours show the outlines of the filament
region, and the [10\%, 90\%] percentiles of the residual values are quoted in
the frames (in units of percent).
}
\label{fig:plot_W}
\end{figure*}

Figure~\ref{fig:plot_W} shows additional fits, where the cloud LOS extent is
optimised as a function of the sky position. These can be compared with the 
fits in the main paper, where the LOS FWHM value was constant over the map
(e.g. Figs.~\ref{fig:plot_residuals_3}-\ref{fig:plot_residuals_4}).  In
Fig.~\ref{fig:plot_W} the errors in the fitted bands (250-500\,$\mu$m) are
getting close to the precision of the observations. The figure quotes the
[10\%, 90\%] percentile ranges for the pixels in the filament region. The
residual maps are shown at 1 arcmin resolution, to highlight the remaining
systematic errors. On the other hand, the FWHM values are $\sim$0.5\,pc or
higher towards the main cores, much larger that the filament POS size and
therefore unlikely to describe the real structure of the cloud.

\begin{figure}
\includegraphics[width=8.8cm]{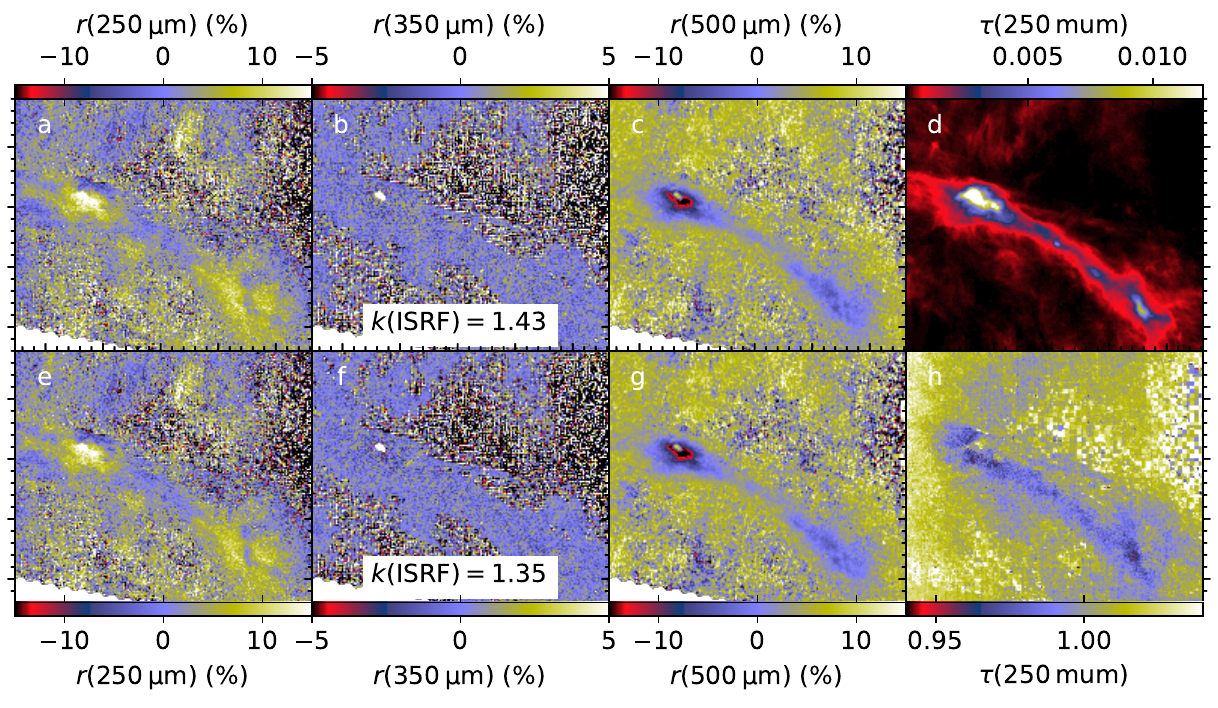}
%
\caption{
Comparison of model fits with different assumptions of filament inclination.
The model assumes $FWHM$=0.5\,pc and $A_{\rm V}=1^{\rm mag}$ and the CMM dust
properties. Frames a-c show the fit residuals for the default model and frames
e-g the residuals when the filament is inclined 45 degrees with respect to
POS. The corresponding scalings of the ISRF levels are given in frames b and
f. Frame d shows the 250\,$\mu$m optical depth for the default model, and
frame h the optical depth in the inclined model relative to the default model.
}
\label{fig:plot_inclined}
\end{figure}

Figure~\ref{fig:plot_inclined} compares results for one cloud model when the
filament is assumed to be either aligned with the POS or inclined by 45
degrees. The effects on the fit quality are practically invisible, and the
effect on the optical depth estimates a couple of percent or less.

\end{appendix}

\end{document}